\documentclass[reqno,centertags, 12pt]{amsart}
\usepackage{amsmath,amsthm,amscd,amssymb}
\usepackage{latexsym}

\newcommand{\bbR}{{\mathbb{R}}}
\newcommand{\bbD}{{\mathbb{D}}}

\newcommand{\bbZ}{{\mathbb{Z}}}
\newcommand{\bbC}{{\mathbb{C}}}

\newcommand{\calE}{{\mathcal E}}

\newcommand{\lb}{\label}
\newcommand{\f}{\frac}
\newcommand{\ul}{\underline}
\newcommand{\ol}{\overline}

\newcommand{\tr}{\text{\rm{Tr}}}

\newcommand{\ess}{\text{\rm{ess}}}
\newcommand{\ac}{\text{\rm{ac}}}

\newcommand{\sing}{\text{\rm{sing}}}

\newcommand{\bi}{\bibitem}

\newcommand{\beq}{\begin{equation}}
\newcommand{\eeq}{\end{equation}}
\newcommand{\ba}{\begin{align}}
\newcommand{\ea}{\end{align}}
\newcommand{\veps}{\varepsilon}

\newcounter{smalllist}
\newenvironment{SL}{\begin{list}{{\rm\roman{smalllist})}}{%
\setlength{\topsep}{0mm}\setlength{\parsep}{0mm}\setlength{\itemsep}{0mm}%
\setlength{\labelwidth}{2em}\setlength{\leftmargin}{2em}\usecounter{smalllist}%
}}{\end{list}}

\DeclareMathOperator{\Ima}{Im}

\allowdisplaybreaks
\numberwithin{equation}{section}

\newtheorem{theorem}{Theorem}[section]
\newtheorem*{t1}{Theorem 1}
\newtheorem*{t2}{Theorem 2}
\newtheorem*{t3}{Theorem 3}
\newtheorem*{t4}{Theorem 4}

\newtheorem{lemma}[theorem]{Lemma}
\newtheorem{corollary}[theorem]{Corollary}
\theoremstyle{definition}

\newtheorem{conjecture}[theorem]{Conjecture}

\theoremstyle{remark}

\newcommand{\abs}[1]{\lvert#1\rvert}

\begin{document}
\title[Sum Rules and the Szeg\H{o} Condition]{Sum Rules and the Szeg\H{o} 
Condition for Orthogonal Polynomials on the Real Line}
\author[B. Simon and A. Zlato\v{s}]{Barry Simon$^1$ and Andrej Zlato\v{s}}

\address{Mathematics 253-37 \\
California Institute of Technology \\
Pasadena, CA 91125 }

\email{bsimon@caltech.edu; andrej@caltech.edu}

\thanks{2000 {\it Mathematics Subject Classification}. Primary: 47B36; Secondary: 42C05}

\thanks{$^1$ Supported in part by NSF grant DMS-9707661. }

\date{May 13, 2002}

\begin{abstract} We study the Case sum rules, especially $C_0$, for general Jacobi matrices. 
We establish situations where the sum rule is valid. Applications include an extension of 
Shohat's theorem to cases with an infinite point spectrum and a proof that if $\lim n 
(a_n -1)=\alpha$ and $\lim nb_n =\beta$ exist and $2\alpha <\abs{\beta}$, then the 
Szeg\H{o} condition fails. 
\end{abstract}

\maketitle

\section{Introduction} \lb{s1}

This paper discusses the relation among three objects well known to be in one-one correspondence: 
nontrivial (i.e., not supported on a finite set) probability measures, $\nu$, of bounded support 
in $\bbR$; orthogonal polynomials associated to geometrically bounded moments; and bounded Jacobi 
matrices. One goes from measure to polynomials via the Gram-Schmidt procedure, from polynomials 
to Jacobi matrices by the three-term recurrence relation, and from Jacobi matrices to measures 
by the spectral theorem.

We will use $J$ to denote the Jacobi matrix ($a_n >0$) 
\begin{equation} \lb{1.1}
J= \begin{pmatrix} 
b_1 & a_1 & 0 & \dots \\
a_1 & b_2 & a_2 & \dots \\
0 & a_2 & b_3 & \dots \\
\dots & \dots & \dots & \dots \\
\end{pmatrix}
\end{equation}
$\nu$ will normally denote the spectral measure of the vector $\delta_1 \in \ell^2 (\bbZ^+)$ and 
$P_n (x)$ the orthonormal polynomials. 

We are interested in $J$'s close to the free Jacobi matrix, $J_0$, with $b_n =0$, $a_n =1$, and 
$d\nu_0 (E)=(2\pi)^{-1} \chi_{[-2,2]} \sqrt{4-E^2}\, dE$. Most often, we will suppose $J-J_0$ is 
compact. That means $\sigma_{\ess}(J)=[-2,2]$ and $J$ has only eigenvalues outside $[-2,2]$, of 
multiplicity one denoted $E_j^\pm$ with $E_1^+ > E_2 ^+ >\cdots > 2$ and $E_1^- < E_2^- < \cdots 
<-2$.

One of the main objects of study here is the Szeg\H{o} integral
\begin{equation} \lb{1.2}
Z(J)=\f{1}{2\pi} \int_{-2}^2 \ln \biggl( \f{\sqrt{4-E^2\,}}{2\pi d\nu_{\ac}/dE}\biggr) 
\f{dE}{\sqrt{4-E^2}}
\end{equation}
The Szeg\H{o} integral is often taken in the literature as 
\[
(2\pi)^{-1} \int_{-2}^2 \ln \biggl( \f{d\nu_{\ac}}{dE}\biggr) \f{dE}{\sqrt{4-E^2\,}}
\]
which differs from $Z(J)$ by a constant and a critical minus sign (so the common condition that 
the Szeg\H{o} integral not be $-\infty$ becomes $Z(J)<\infty$ in our normalization). There is an 
enormous literature discussing when $Z(J)<\infty$  holds (see, e.g., \cite{AI,Case,DN,Gon,Nev1, 
NevAMS,Nik,PY,Sh,Sz20}). It can be shown by Jensen's inequality that $Z(J)\geq -\f12 \ln(2)$ 
so the integral can only diverge to $+\infty$. 

We will focus here on various sum rules that are valid. One of our main results is the 
following:

\begin{t1}\lb{T1} Suppose 
\begin{equation} \lb{1.3}
A_0 (J) =\lim_{N\to\infty}\, \biggl( -\sum_{n=1}^N \ln (a_n)\biggr) 
\end{equation}
exists {\rm{(}}although it may be $+\infty$ or $-\infty${\rm{)}}. Consider the additional 
quantities $Z(J)$ given by \eqref{1.2} and 
\begin{equation} \lb{1.4}
\calE_0 (J) =\sum_\pm \sum_j \ln \biggl[ \tfrac12 \biggl(\abs{E_j^\pm} + 
\sqrt{(E_j^\pm)^2 -4\,}\biggr)\biggr]
\end{equation}
If any two of the three quantities $A_0 (J)$, $\calE_0 (J)$, and $Z(J)$ are finite, then 
all three are, and 
\begin{equation} \lb{1.5}
Z(J) =A_0 (J) + \calE_0 (J)
\end{equation}
\end{t1}

\smallskip
{\it Remarks.} 1. It is not hard to see that $\calE_0 (J)<\infty$ if and only if 
\begin{equation} \lb{1.5new}
\sum_\pm \sum_j \sqrt{(E_j^\pm)^2 -4} < \infty
\end{equation}

\smallskip
2. The full theorem (Theorem~\ref{T4.1}) does not require the limit \eqref{1.3} 
to exist, but is more complicated to state in that case.

\smallskip
3. If the three quantities are finite, many additional sum rules hold.

\smallskip
4. This is what Killip-Simon \cite{KS} call the $C_0$ sum rule. 

\smallskip
5. Peherstorfer-Yuditskii \cite{PY} (see their remark after Lemma~2.1) prove that if $Z(J)<\infty$, 
$\calE_0(J)=\infty$, then the limit in \eqref{1.3} is also infinite. 

\smallskip
Theorem~1 is an analog for the real line of a seventy-year old theorem for orthogonal 
polynomials on the unit circle:
\begin{equation} \lb{1.6}
\f{1}{2\pi} \int_0^{2\pi} \ln \biggl( \f{d\nu_{ac}}{d\theta}\biggr)\, d\theta = 
\sum_{n=0}^\infty \ln (1-\abs{\alpha_j}^2)
\end{equation}
where $\{\alpha_j\}_{j=1}^\infty$ are the Verblunsky coefficients (also called reflection, 
Geronimus, Schur, or Szeg\H{o} coefficients) of $\nu$. This result was first proven by 
Verblunsky \cite{Ve} in 1935, although it is closely related to Szeg\H{o}'s 1920 paper \cite{Sz20}.

For $J$'s with $J-J_0$ finite rank (and perhaps even with $\sum_{n=1}^\infty n(\abs{a_n-1} 
+\abs{b_n})<\infty$), the sum rule \eqref{1.5} is due to Case \cite{Case}. Recently, Killip-Simon 
\cite{KS} showed how to exploit these sum rules as a spectral tool (motivated in turn by 
work on Schr\"odinger operators by Deift-Killip \cite{DK} and Denissov \cite{Den}). In particular, 
Killip-Simon emphasized the importance in proving sum rules on as large a class of $J$'s 
as possible. 

One application we will make of Theorem~\ref{T1} and related ideas is to prove the following 
($\equiv$ Theorem~\ref{T5.2}):

\begin{t2}\lb{T2} Suppose $\sigma_{\ess} (J)\subset [-2,2]$ and \eqref{1.5new} holds. Then 
$Z(J)<\infty$ if and only if 
\begin{equation} \lb{1.7}
\liminf_N \biggl( -\sum_{n=1}^N \ln (a_n)\biggr) <\infty
\end{equation}
Moreover, if these conditions hold, then 
\begin{SL} 
\item[{\rm{(i)}}] The limit $A_0 (J)$ in \eqref{1.3} exists and is finite. 
\item[{\rm{(ii)}}] $\lim_{N\to\infty} \sum_{n=1}^N b_n$ exists and is finite. 
\item[{\rm{(iii)}}] 
\begin{equation} \lb{1.7a} 
\sum_{n=1}^\infty (a_n -1)^2 + \sum_{n=1}^\infty b_n^2 <\infty
\end{equation}
\end{SL}
\end{t2}

Results of this genre when it is assumed that $\sigma(J)=[-2,2]$ go back to Shohat \cite{Sh} with 
important contributions by Nevai \cite{NevAMS}. The precise form is from Killip-Simon \cite{KS}. 
Nikishin \cite{Nik} showed how to extend this to Jacobi matrices with finitely many eigenvalues. 
Peherstorfer-Yuditskii \cite{PY} proved $Z(J)<\infty$ implies (i) under the condition $\calE_0 (J) 
<\infty$, allowing an infinity of eigenvalues for the first time. Our result cannot extend 
to situations with $\calE_0 (J)=\infty$ since Theorem~\ref{T1} says if (i) holds and $Z(J)<
\infty$, then $\calE_0 (J)<\infty$.

We will highlight one other result we will prove later (Corollary~6.3).

\begin{t3}\lb{T3} Let $a_n, b_n$ be Jacobi matrix parameters so that 
\begin{equation} \lb{1.8}
\lim_{n\to\infty}\, n(a_n -1)=\alpha \qquad \lim_{n\to\infty} \, nb_n =\beta
\end{equation}
exist and are finite. Suppose that 
\begin{equation} \lb{1.9}
\abs{\beta} >2\alpha 
\end{equation}
Then $Z(J)=\infty$.
\end{t3}

{\it Remark.} In particular, if $\alpha <0$, \eqref{1.9} always holds. \eqref{1.9} describes 
three-quarters of the $(2\alpha,\beta)$ plane.

\smallskip
In Section~\ref{s6}, we will discuss the background for this result, and describe results of 
Zlato\v{s} \cite{Zl} that show if $\abs{\beta} \leq 2\alpha$ and one has additional information 
on the approach to the limit \eqref{1.8}, then $Z(J)<\infty$. Thus Theorem~3 captures the precise 
region where one has \eqref{1.8} and one can hope to prove $Z(J)=\infty$.

Theorem~3 will actually follow from a more general result (see Theorem~\ref{T4.4}, \ref{T6.1}, 
and \ref{new}).

\begin{t4} \lb{T4} Suppose \eqref{1.7a} holds and that either $\limsup (-\sum_{j=1}^n (a_j -1 
+ \f12 b_j))=\infty$ or $\limsup (-\sum_{j=1}^n (a_j -1 - \f12 b_j)) =\infty$. Then $Z(J)
=\infty$. 
\end{t4}

The main technique in this paper exploits the $m$-function, the Borel transform of the measure, $\nu$: 
\begin{equation} \lb{1.10}
m_\nu (E) =\int \f{d\nu (x)}{x-E}
\end{equation}
Since $\nu$ is supported on $[-2,2]$ plus the set of points $\{E_j^\pm\}$, we can write 
\begin{equation} \lb{1.11}
m_\nu (E) =\sum_\pm \sum_j \f{\nu(\{E_j^\pm\})}{E_j^\pm -E} + \int_{-2}^2 \f{d\nu(x)}{x-E}
\end{equation}

It is useful to transfer everything to the unit circle, using the fact that $z\mapsto E=z+z^{-1}$ 
maps $\bbD =\{z\mid\,\abs{z}<1\}$ onto the cut plane $\bbC\backslash [-2,2]$. Thus we can define 
for $\abs{z}<1$ 
\begin{equation} \lb{1.12}
M(z) =-m_\nu (z+z^{-1})
\end{equation}
The minus sign is picked so $\Ima M(z)>0$ if $\Ima z>0$. We use $M(z;J)$ when we want to make 
the $J$-dependence explicit. 

In \eqref{1.11}, we translate the pole term directly. Define $\beta_j$ so 
\begin{equation} \lb{1.13}
E_j^\pm =\beta_j^\pm + (\beta_j^\pm)^{-1}
\end{equation}
with $\abs{\beta_j^\pm} >1$. We sometimes drop the explicit $\pm$ symbol and count the 
$\beta_j$'s in one set. We define a signed measure $d\mu^\#$ on $(0,2\pi)$ by $\Ima M
(re^{i\theta})\to d\mu^\#$ weakly as $r\uparrow 1$. $\mu^\#$ is positive on $(0,\pi)$ and 
negative on $(\pi, 2\pi)$. Thus \eqref{1.11} implies 
\begin{equation} \lb{1.14}
\Ima M(z)= \Ima \sum_j \f{\mu (\{\beta_j^{-1}\})}{z+z^{-1} - (\beta_j + \beta_j^{-1})} + \f{1}{2\pi} 
\int_0^{2\pi} P(z, e^{i\varphi})\, d\mu^\# (\varphi)
\end{equation}
where we use $\mu(\{\beta_j^{-1}\})$ for the weights $\nu(\{E_j\})$ and $P(z,w)$, with 
$\abs{z}\leq 1$, $\abs{w} =1$, is the Poisson kernel 
\begin{equation} \lb{1.15}
P(z, w) =\f{1-\abs{z}^2}{\abs{z-w}^2}
\end{equation}
or 
\begin{equation} \lb{1.16}
P(re^{i\theta}, e^{i\varphi})\equiv P_r (\theta-\varphi) = 
\f{1-r^2}{1+r^2 -2r\cos (\theta-\varphi)} 
\end{equation}
We note now that since $\mu(\{\beta_j^{-1}\})$ are point mass of a probability measure, we have 
\[
\sum_j \mu(\{\beta_j^{-1}\}) \leq 1
\]

It is useful to use the fact that $\mu^\#$ is odd under reflection to rewrite \eqref{1.14} in the 
form 
\begin{equation} \lb{1.17}
\Ima M(re^{i\theta}) = \Ima \sum_j \f{\mu(\{\beta_j^{-1}\})}
{re^{i\theta} + r^{-1} e^{-i\theta} - (\beta_j +\beta_j^{-1})} + \f{1}{2\pi} 
\int_0^\pi D_r (\theta, \varphi)\, d\mu(\varphi)
\end{equation}
where
\begin{equation} \lb{1.18}
D_r (\theta, \varphi) = P_r (\theta, \varphi) - P_r (\theta, -\varphi)
\end{equation}
This is because $M(z) =\ol{M(\bar z)}$, so that $\mu^\#\restriction [-\pi, 0]$ is a reflection 
across $\bbR$ of $\mu\equiv \mu^\# \restriction [0,\pi]$.

Note that not only does $\Ima M(re^{i\theta}) d\theta$ converge weakly to $\mu^\#$, but by 
general principles \cite{Rudin}, $\lim_{r\uparrow 1} M(re^{i\theta})\equiv M(e^{i\theta})$ exists 
for a.e.~$\theta$ and $d\mu_{\ac}^\# =\Ima M(e^{i\theta}) d\theta$. 

Section~\ref{s2}, the technical core of the paper, proves some convergence results about integrals 
of $\ln [\Ima M(re^{i\theta})]$. It is precisely such integrals that arise in Section~\ref{s3} where,  
following Killip-Simon \cite{KS}, we use the well-known 
\[
-m(z;J)^{-1} =z-b_1 + a_1^2 m(z;J^{(1)})
\]
where $J^{(1)}$ is $J$ with the top row and leftmost column removed. We will be able to prove 
sum rules that compare $J$ and $J^{(1)}$. In Section~\ref{s4}, we will then list various sum rules, 
including Theorems~1 and 4. Section~\ref{s5} proves Theorem~2 and Section~\ref{s6} discusses 
Coulomb Jacobi matrices ($J-J_0$ decays as $n^{-1}$) and Theorem~3 in particular.

It is a pleasure to thank Mourad Ismail, Rowan Killip, and Paul Nevai for useful discussions.

\bigskip
\section{Continuity of Integrals of $\ln (\Ima M)$} \lb{s2}

In this section, we will prove a general continuity result about boundary values of interest 
for $M$-functions of the type defined in \eqref{1.16}. We will consider suitable weight functions, 
$w(\varphi)$, on $[0,\pi]$, of which the examples of most interest are $w(\varphi)=\sin^k 
(\varphi)$, $k=0$ or $2$. Our goal is to prove that 
\begin{equation} \lb{2.1} 
\lim_{r\uparrow 1} \int \ln [\Ima M(re^{i\varphi})]\, w(\varphi)\, d\varphi =
\int \ln [\Ima M(e^{i\varphi})] \, w(\varphi)\, d\varphi
\end{equation}
and that the convergence is in $L^1$ if the integral on the right is finite. All integrals in 
this section are from $0$ to $\pi$ if not indicated otherwise. We define 
\begin{equation} \lb{2.2} 
d(\varphi)\equiv \min (\varphi, \pi-\varphi)
\end{equation}
and we suppose that  
\begin{equation} \lb{2.3} 
0 \leq w(\varphi) \leq C_1 \, d(\varphi)^{-1+\alpha}
\end{equation}
for some $C_1, \alpha >0$ and that $w$ is $C^1$ with 
\begin{equation} \lb{2.4} 
\abs{w'(\varphi)w(\varphi)^{-1}}\leq C_2 \, d(\varphi)^{-\beta}
\end{equation}
for $C_2,\beta>0$. For weights of interest, one can take $\alpha=\beta=1$.

\smallskip
{\it Remarks.} 1. For the applications in mind, we are only interested in allowing 
``singularities" (i.e., $w$ vanishing or going to infinity) at $0$ or $\pi$, but all results 
hold with unchanged proofs if $d(\varphi) \equiv \min \{\abs{\varphi - \varphi_j}\}$ for any finite 
set $\{\varphi_j\}$. For example, $w(\varphi)=\sin^2 (m\varphi)$ as in \cite{LNS} is fine.

\smallskip
2. Note that by \eqref{2.3}, $\int_0^\pi w(\varphi)\, d\varphi <\infty$.

\smallskip

The main technical result we will need is:

\begin{theorem} \lb{T2.1} Let $M$ be a function with a representation of the form \eqref{1.17} 
and let $w$ be a weight function obeying \eqref{2.3} and \eqref{2.4}. Then \eqref{2.1} holds. 
Moreover, if 
\begin{equation} \lb{2.4a} 
\int \ln [\Ima M(e^{i\varphi})] w(\varphi)\, d\varphi > -\infty
\end{equation}
{\rm{(}}it is never $+\infty${\rm{)}}, then 
\begin{equation} \lb{2.5} 
\lim_{r\uparrow 1}\, \int \bigl| \ln [\Ima M(re^{i\varphi})] -\ln [\Ima M(e^{i\varphi})]\bigr| \, 
w(\varphi)\, d\varphi =0
\end{equation}
\end{theorem}

Let $\ln_\pm$ be defined by 
\[
\ln_\pm (y) =\max (0, \pm\ln (y))
\]
so 
\begin{align*}
\ln (y) &= \ln_+ (y) -\ln_- (y) \\
\abs{\ln(y)} &=\ln_+ (y) + \ln_- (y) 
\end{align*}

We will prove Theorem~\ref{T2.1} by proving 

\begin{theorem} \lb{T2.2} For any $a>0$ and $p<\infty$, $\ln_+ [\Ima (M(e^{i\varphi}))/a]\in L^p 
((0,\pi), w(\varphi) d\varphi)$, and 
\begin{equation} \lb{2.6} 
\lim_{r\uparrow 1}\, \int \biggl| \ln_+ \bigg( \f{\Ima M(re^{i\varphi})}{a}\biggr) - 
\ln_+ \biggl( \f{\Ima M(e^{i\varphi})}{a}\biggr)\biggr|^p\,  w(\varphi)\, d\varphi =0
\end{equation}
\end{theorem}

\begin{theorem}\lb{T2.3} For any $a>0$, we have 
\begin{equation} \lb{2.7} 
\lim_{r\uparrow 1} \int \ln_- \biggl( \f{\Ima M(re^{i\varphi})}{a}\biggr) w(\varphi)\, d\varphi 
=\int \ln_- \biggl( \f{\Ima M(e^{i\varphi})}{a}\biggr) w(\varphi)\, d\varphi
\end{equation}
\end{theorem}

\begin{proof}[Proof of Theorem~\ref{T2.1} given Theorems~\ref{T2.2} and \ref{T2.3}] By Fatou's 
lemma and the fact that for a.e.~$\varphi$, $\Ima M(re^{i\varphi}) \to \Ima M(e^{i\varphi})$, 
we have 
\begin{equation} \lb{2.8} 
\liminf_{r\uparrow 1}\,\int \ln_- [\Ima M(re^{i\varphi})]\, w(\varphi)\, d\varphi \geq 
\int \ln_- [\Ima M(e^{i\varphi})]\, w(\varphi)\, d\varphi
\end{equation}
Since Theorem~\ref{T2.2} says that $\sup_{0<r\leq 1} \int \ln_+ [\Ima M(re^{i\varphi})]  
w(\varphi)\, d\varphi <\infty$, it follows that if $\int \ln_- [\Ima M(e^{i\varphi})] 
w(\varphi)\, d\varphi = \infty$, then \eqref{2.1} holds.

If \eqref{2.4a} holds, then 
\[
\lim_{a\downarrow 0}\, \int \ln_- \biggl[ \f{\Ima M(e^{i\varphi})}{a}\biggr] w(\varphi) \,  
d\varphi =0
\]
since $\ln_- (y/a)$ is monotone decreasing to $0$ as $a$ decreases. Given $\veps$, first find 
$a$ so 
\[
\int \ln_- \biggl[ \f{\Ima M(e^{i\varphi})}{a}\biggr] w(\varphi)\, d\varphi < \f{\veps}{3}
\]
and then, by \eqref{2.7}, $r_1<1$ so for $r_1<r<1$, 
\[
\int \ln_- \biggl[ \f{\Ima M(re^{i\varphi})}{a}\biggr] w(\varphi)\, d\varphi <\f{\veps}{3}
\]
By \eqref{2.6}, find $r_2 <1$, so for $r_2 < r < 1$, 
\[
\int \biggl| \ln_+ \biggl[ \f{\Ima M (re^{i\varphi})}{a}\biggr] - 
\ln_+ \biggl[ \f{\Ima M(e^{i\varphi})}{a}\biggr] \biggr|\, w(\varphi)\, d\varphi < \f{\veps}{3}
\]

Writing 
\[
\abs{\ln (\alpha) -\ln (\beta)} \leq \biggl| \ln_+ \biggl( \f{\alpha}{a}\biggr) - 
\ln_+ \biggl( \f{\beta}{a}\biggr) \biggr| + \ln_- \biggl( \f{\alpha}{a}\biggr) + 
\ln_- \biggl( \f{\beta}{a}\biggr)
\]
we see that if $\max (r_1, r_2) < r<1$, then 
\[
\int \bigl| \ln [\Ima M(re^{i\varphi})] -\ln [\Ima M(e^{i\varphi})]\bigr|\, w(\varphi)\, 
d\varphi <\veps
\]
so \eqref{2.5} holds. 
\end{proof}

We will prove Theorem~\ref{T2.2} by using the dominated convergence theorem and standard maximal 
function techniques. Given the measure $\mu$ on $(0,\pi)$, we define $\tilde\mu$ to be the sum 
of $\mu$ and its reflection on $(\pi, 2\pi)$, and its maximal function by 
\[
\mu^* (x) =\sup_{0<a<\pi}\, \f{\tilde\mu(x-a, x+a)}{2a} 
\]
The Hardy-Littlewood maximal inequality for measures (see Rudin \cite{Rudin}) says that   
\begin{equation} \lb{2.9} 
\abs{\{x\mid \mu^*(x) >\lambda\}}\leq \f{3\tilde\mu(0,\pi)}{\lambda}
\end{equation}

\begin{lemma}\lb{L2.4} Let $M$ be an $M$-function based on a measure $\mu$ on $[0,\pi]$ and 
weights at the poles at $(\beta_j^\pm)^{-1}$, and let $\alpha$ be a sum of the weights of the 
poles. Then for $0<r<1$, 
\begin{equation} \lb{2.10} 
\Ima M (re^{i\theta}) \leq \mu^* (\theta) +\alpha r^{-1} [\sin(\theta)]^{-2}
\end{equation}
\end{lemma}

\begin{proof} Since $D_r (\theta, \varphi)\leq P_r (\theta, \varphi)$ and $P_r$ is a convolution 
operator with a positive even decreasing function of $\varphi$ on $[0,\pi]$ with $\int_0^{2\pi} 
P_r (\varphi) \, d\varphi/2\pi =1$, we have, by standard calculations, that 
\[
\int_0^\pi D_r (\theta, \varphi) \, \f{d\mu(\varphi)}{2\pi} \leq \mu^* (\theta)
\]

On the other hand, for $\beta\geq 1$, 
\[
\biggl| \f{1}{z+z^{-1} -\beta -\beta^{-1}}\biggr| = \biggl| \f{z}{(z-\beta)(z-\beta^{-1})}
\biggr| \leq \f{\abs{z}}{\abs{\Ima z}^2} \leq \f{1}{r\sin^2 \theta}
\]
if $z=re^{i\theta}$, so summing the pole term shows, 
\[
\Ima \sum_i \f{\mu(\{\beta_i^{-1}\})}{z+z^{-1} -\beta_i - \beta_i^{-1}} \leq 
\f{\sum_i \mu(\{\beta_i^{-1}\})}{r\sin^2 \theta}
\]
\end{proof}

\begin{proof}[Proof of Theorem~\ref{T2.2}] Let 
\[
f_1 (\theta) =\mu^*(\theta) \qquad f_2 (\theta) =2 [\sin(\theta)]^{-2}
\]
For a.e.~$\theta$, $\ln_+ [(\Ima M(re^{i\theta}))/a] \to \ln_+ [(\Ima M(e^{i\theta}))/a]$. By 
\eqref{2.10} for all $\f12 < r<1$, $\ln_+ [(\Ima M(re^{i\theta}))/a]\leq \ln_+ [(f_1 (\theta) + 
f_2(\theta))/a]$. Thus if we prove that for all $p<\infty$, 
\[
\int \biggl| \ln_+ \biggl( \f{f_1 + f_2}{a}\biggr)\biggr|^p w(\varphi)\, d\varphi <\infty
\]
we obtain \eqref{2.6} by the dominated convergence theorem. Since 
\[
\abs{\ln_+(x)}^p \leq C(p,q) \abs{x}^q
\]
for any $p<\infty$, $q>0$, and suitable $C(p,q)$ and 
\[
\abs{x+y}^q \leq 2^q \abs{x}^q + 2^q \abs{y}^q
\]
it suffices to find some $q>0$, so 
\[
\int (\abs{f_1 (\varphi)}^q + \abs{f_2 (\varphi)}^q)\, w(\varphi)\, d\varphi <\infty
\]

Since for $v^{-1} + t^{-1} =1$, 
\[
\int \abs{f_1 (\varphi)}^q w(\varphi)\, d\varphi \leq \biggl( \int \abs{f_1 (\varphi)}^{qv} 
\, d\varphi\biggr)^{1/v} \biggl( \int \abs{w(\varphi)}^t \,d\varphi\biggr)^{1/t}
\]
and $w(\varphi)\in L^t$ for some $t>1$ by \eqref{2.3}, it suffices to find some $s>0$ with 
\begin{equation} \lb{2.10x}
\int (\abs{f_1 (\varphi)}^s + \abs{f_2 (\varphi)}^s)\, d\varphi <\infty 
\end{equation} 
By \eqref{2.9}, $\int \abs{f_1 (\varphi)}^s\, d\varphi <\infty$ if $s<1$ and clearly, $\int 
\abs{f_2 (\varphi)}^s \, d\varphi <\infty$ if $s<\f12$. 
\end{proof}

As a preliminary to the proof of Theorem~\ref{T2.3}, we need

\begin{lemma}\lb{L2.5} Let $w$ obey \eqref{2.4}. Let $0<\varphi_0<\pi$ and let $\varphi_1,  
\varphi_2 \in [0,\pi]$ obey 
\begin{alignat}{2} 
&\text{{\rm{(a)}}} \qquad && d(\varphi_1) \geq d(\varphi_0),\quad d(\varphi_2)\geq d(\varphi_0) 
\lb{2.11} \\
&\text{{\rm{(b)}}} \qquad && \abs{\varphi_1 -\varphi_2} \leq d(\varphi_0)^\beta \lb{2.12}
\end{alignat}
Then for $C_3 = C_2 \exp (C_2)$, 
\begin{equation} \lb{2.13} 
\biggl| \f{w(\varphi_1)}{w(\varphi_2)} -1 \biggr| \leq C_3 \abs{\varphi_1 -\varphi_2}\, 
d(\varphi_0)^{-\beta}
\end{equation}
\end{lemma}

\begin{proof} 
\begin{align} 
\biggl| \f{w(\varphi_1)}{w(\varphi_2)} -1 \biggr| &= \biggl| \exp \biggl(\, 
\int_{\varphi_1}^{\varphi_2} \f{w'(\eta)}{w(\eta)}\, d\eta\biggr) -1\biggr| \notag \\
&\leq \abs{\exp (C_2 \abs{\varphi_2 -\varphi_1}\, d(\varphi_0)^{-\beta}) -1 } \lb{2.14} 
\end{align}
by \eqref{2.4} and \eqref{2.11}. But $\abs{e^x -1} \leq e^{\abs{x}}\abs{x}$, so by \eqref{2.12}, 
\[
\biggl| \f{w(\varphi_1)}{w(\varphi_2)} -1\biggr| \leq C_2 \exp (C_2) \abs{\varphi_1 -\varphi_2} 
\, d(\varphi_0)^{-\beta}
\]
which is \eqref{2.13}. 
\end{proof}

We will also need the following pair of lemmas:

\begin{lemma}\lb{L2.6} Let $0<\eta<\theta<\pi-\eta$ and 
\[
N_r (\theta, \eta) =\int_{\theta-\eta}^{\theta+\eta} D_r (\theta,\varphi)\, \f{d\varphi}{2\pi} 
\]
Then 
\begin{equation} \lb{2.15} 
0\leq [1-N_r (\theta,\eta)] \leq \f{2(1-r)}{r\sin^2\theta} + \f{1-r}{r\sin^2\eta}
\end{equation}
\end{lemma}

\begin{proof} We have 
\[
1=\int_0^{2\pi} P_r (\theta, \varphi)\, \f{d\varphi}{2\pi}
\]
so since $D_r \leq P_r$, $N_r \leq 1$ and 
\begin{equation} \lb{2.16} 
1-N_r (\theta, \eta) \leq \f{2}{2\pi} \int_{-\pi}^0 P_r (\theta, \varphi)\, d\varphi + 
\f{1}{2\pi} \int_{\substack{ \varphi\in [0,\pi] \\ \abs{\theta-\varphi}\geq\eta}} 
P_r (\theta,\varphi)\, d\varphi
\end{equation}
Now 
\begin{align}
P_r (\theta, \varphi) &= \f{1-r^2}{(1-r)^2 + 4r\sin^2 [\f12(\theta-\varphi)]} \lb{2.17} \\
&\leq \f{2(1-r)}{4r\sin^2 [\f12(\theta-\varphi)]} \notag \\
&\leq \f{2(1-r)}{r\sin^2 (\theta-\varphi)} \notag  
\end{align}

The first integrand in \eqref{2.16} is thus bounded by $2r^{-1} (1-r)[\sin^2 (\theta)]^{-1}$ 
and the second by $2r^{-1} (1-r) [\sin^2 (\eta)]^{-1}$, so \eqref{2.15} is immediate. 
\end{proof}

\begin{lemma} \lb{L2.7} If $\int_0^\pi \Ima M(e^{i\theta})\, d\theta\neq 0$, then for $\theta\in 
[0,\pi]$, $r\in (\f12, 1)$, 
\begin{equation} \lb{2.18} 
\Ima M (re^{i\theta}) \geq c (r^{-1} -r) \sin\theta
\end{equation}
\end{lemma}

\begin{proof} In terms of the real line $m$ function, for $E_2 >0$, $E_1$ real,  
\begin{equation} \lb{2.19} 
\Ima [-m (E_1 -i E_2)] \geq \f{E_2}{\pi} \int_{-2}^2 \f{\Ima m (E)\, dE}
{(E_1-E)^2 + E_2^2}
\end{equation}
since we have dropped the positive contributions of $\nu_{\sing}$ to $\Ima (-m)$. 
Now if $z=re^{i\theta}$, 
\[
M(z) =-m(E_1 -iE_2)
\]
where $z+z^{-1} =E_1 -iE_2$ or $E_1 =(r+r^{-1}) \cos \theta$, $E_2 =(r^{-1}-r)\sin \theta$. 
If $r>\f12$, then $\abs{E_1} \leq \f{5}{2}$, $\abs{E_2} \leq \f32$, and in \eqref{2.19}, 
$\abs{E}\leq 2$. Thus 
\[
\Ima M(z) \geq c E_2 (z)
\]
which is \eqref{2.18}. 
\end{proof}

\begin{proof}[Proof of Theorem~\ref{T2.3}] Since $\ln_-$ is a decreasing function, to get upper 
bounds on $\ln_- [\Ima M(re^{i\theta})/a]$, we can use a lower bound on $\Ima M$. The elementary 
bound  
\begin{equation} \lb{2.20} 
\ln_- (ab)\leq \ln_- (a) + \ln_-(b)
\end{equation}
will be useful.

As already noted, Fatou's lemma implies the $\liminf$ of the left side of \eqref{2.7} is bounded 
from below by the right side, so it suffices to prove that 
\begin{equation} \lb{2.21x} 
\limsup_{r\uparrow 1}\, \int_0^\pi \ln_- \biggl( \f{\Ima M(re^{i\varphi})}{a}\biggr) w(\varphi)\, 
d\varphi \leq \int_0^\pi \ln_- \biggl( \f{\Ima M(e^{i\varphi})}{a}\biggr) w(\varphi)\, d\varphi
\end{equation}

Pick $\gamma$ and $\kappa$ so $0<\max (\beta, 1)\gamma < \kappa < \f12$ and let $\theta_0(r) =
(1-r)^\gamma$, $\eta(r)=(1-r)^\kappa$. We will bound $\Ima M(re^{i\theta})$ from below for 
$d(\theta) \leq \theta_0 (r)$ using \eqref{2.18}, and for $d(\theta)\geq \theta_0(r)$, we will 
use the Poisson integral for the region $\abs{\varphi-\theta}\leq \eta(r)$. 

By \eqref{2.18} and \eqref{2.3}, 
\[
\int_{d(\varphi) \leq \theta_0 (r)} \ln_- \biggl( \f{\Ima M(re^{i\varphi})}{a}\biggr) 
w(\varphi) \, d\varphi \leq C_a \theta_0^\alpha [\ln_- (r^{-1}-r) + \ln_- \theta_0]
\]
which goes to zero as $r\uparrow 1$ for any $a$. So suppose $d(\theta)>\theta_0$. Write 
\begin{align}
\Ima M(re^{i\theta}) &\geq \int_{\theta-\eta(r)}^{\theta + \eta(r)} D_r (\theta, \varphi) 
\Ima M(e^{i\varphi}) \, \f{d\varphi}{2\pi} \notag \\
&= N_r (\theta, \eta) \int_{\theta-\eta(r)}^{\theta+\eta(r)} \f{D_r (\theta, \varphi)}
{2\pi N_r (\theta, \eta)}\, \Ima M(e^{i\varphi})\, d\varphi \lb{2.21}
\end{align}

For later purposes, note that for $d(\theta) >\theta_0$, \eqref{2.15} implies 
\begin{equation} \lb{2.22} 
0\leq 1 -N_r (\theta, \eta) \leq C(1-r)^{1-2\kappa}
\end{equation}
which goes to zero since $\kappa <\f12$. Using \eqref{2.21} and \eqref{2.20}, we bound $\ln_- 
[\Ima M(re^{i\theta})/a]$ as two $\ln_-$'s. Since $\ln_-$ is convex and $D_r (\theta,\varphi)/ 
2\pi N_r (\theta, \eta)$ $\chi_{(\theta-\eta, \theta+\eta)}(\varphi)\, d\varphi$ is a probability 
measure, we can use Jensen's inequality to see that 
\begin{equation} \lb{2.23} 
\begin{split}
w(\theta) &\ln_- \biggl[\f{\Ima M(re^{i\theta})}{a}\biggr] \\
& \leq w(\theta) \ln_- [N_r (\theta, \eta)] + \int_{\theta- \eta(r)}^{\theta + \eta(r)} 
\f{w(\theta)}{w(\varphi)}\, \f{D_r (\theta, \varphi)}{N_r (\theta, \eta)}\, 
w(\varphi) \ln_- \biggl[\f{\Ima M(e^{i\varphi})}{a}\biggr] \, \f{d\varphi}{2\pi}
\end{split}
\end{equation}

In the first term for the $\theta$'s with $d(\theta)\geq \theta_0 (r)$, $N_r$ obeys \eqref{2.22} 
so 
\begin{equation} \lb{2.24} 
\int_{d(\theta) \geq \theta_0(r)} w(\theta) \ln_- [N_r (\theta,\eta)] \, d\theta
= O((1-r)^{1-2\kappa})\to 0
\end{equation}
In the second term, note that for the $\theta$'s in question, $N_r (\theta, \eta)^{-1} -1 
=O((1-r)^{1-2\kappa})$ and by \eqref{2.13}, $w(\theta)/w(\varphi) -1 =
O((1-r)^{\kappa -\beta\gamma})$. Since $D_r (\theta, \varphi) \leq P_r (\theta, \varphi)$, 
we thus have 
\begin{multline}\lb{2.25} 
\int_{d(\theta)\geq\theta_0} \ln_- \biggl[\f{\Ima M(re^{i\theta})}{a}\biggr]w(\theta)\, d\theta \\
 \leq O((1-r)^{1-2\kappa}) + [1+ O((1-r)^{1-2\kappa})] [1 + O((1-r)^{\kappa-\beta\gamma})] \\
 \int_{\substack{ d(\theta)\geq \theta_0 \\ \abs{\varphi-\theta}\leq\eta}} P_r (\theta, \varphi) 
w(\varphi) \ln_- \biggl[ \f{\Ima M(e^{i\varphi})}{a}\biggr]\, d\varphi \, \f{d\theta}{2\pi}
\end{multline}

Since the integrand is positive, we can extend it to $\{(\theta, \varphi)\mid\theta\in [0,2\pi], 
\varphi\in [0,\pi]\}$ and do the $\theta$ integration using $\int P_r (\theta, \varphi) 
d\theta/2\pi =1$. The result is \eqref{2.21x}. 
\end{proof}

This concludes the proof of Theorem~\ref{T2.1}. By going through the proof, one easily sees 
that

\begin{theorem}\lb{T2.8} Theorem~\ref{T2.1} remains true if in \eqref{2.1} and \eqref{2.5}, 
$\ln [\Ima M(re^{i\varphi})]$ is replaced by $\ln [g(r)\sin\varphi + \Ima M (re^{i\varphi})]$ 
where $g(r)\geq 0$ and $g(r)\to 0$ as $r\uparrow 1$. 
\end{theorem}

\begin{proof} In the $\ln_+$ bounds, we get an extra $[\sup_{\f12 < r < 1} g(r)]\sin\theta$ 
in $f_2 (\theta)$. Since we still have pointwise convergence, we easily get the analog of 
Theorem~\ref{T2.2}. In the proof of Theorem~\ref{T2.3}, Fatou is unchanged since $g(r)\to 0$,  
and since 
\[
\ln_- (g(r)\sin\varphi + \Ima M(re^{i\varphi}))\leq \ln_- (\Ima M(re^{i\varphi}))
\]
the $\limsup$ bound has an unchanged proof. 
\end{proof}

\bigskip
\section{The Step-by-Step Sum Rules} \lb{s3} 

We will call $J$ a BW matrix (for Blumenthal-Weyl) if $J$ is a bounded Jacobi matrix with 
$\sigma_{\ess} (J)=[-2,2]$, for example, if $J-J_0$ is compact. Let  $J^{(n)}$ be the matrix 
resulting from removing the first $n$ rows and columns. Let $\{E_j^\pm(J)\}_{j=1}^\infty$ 
be the eigenvalues of $J$ above/below $\pm 2$, ordered by $\pm E_1^\pm \geq \pm E_2^\pm 
\geq \cdots$ with $E_j^\pm (J)$ defined to be $\pm 2$ if there are only finitely many 
eigenvalues $k<j$ above/below $\pm 2$. Then by the min-max principle,  
\begin{equation} \lb{3.1}
\pm E_{j+n}^\pm (J) \leq \pm E_j^\pm (J^{(n)})\leq \pm E_j^\pm (J)
\end{equation} 
We have $\lim_{j\to\infty} E_j^\pm (J) =\pm 2$ if $J$ is a BW matrix. 

It follows by the convergence of sums of alternating series that if $f$ is even or odd and 
monotone on $[2,\infty)$ with $f(2)=0$, then 
\begin{equation} \lb{3.2}
\lim_{N\to\infty} \sum_\pm \, \sum_{j=1}^N \,  [ f(E_j^\pm (J)) - f(E_j^\pm (J^{(n)}))] \equiv 
\delta f_n (J)
\end{equation}
exists and is finite. If $\beta_j^\pm$ is defined by $E_j^\pm = \beta_j^\pm + 
(\beta_j^\pm)^{-1}$ with $\abs{\beta_j}>1$, we define $X_\ell^{(n)}(J)$ as $\delta f_n (J)$ for 
\begin{equation} \lb{3.3}
f(E) = \begin{cases} 
\ln \abs{\beta} & \ell=0 \\
-\f{1}{\ell} [\beta^\ell - \beta^{-\ell} ] & \ell = 1,2 \dots 
\end{cases}
\end{equation}

In addition, we will need 
\begin{equation} \lb{3.4}
\zeta_\ell^{(n)}(J) = \begin{cases} 
-\sum_{j=1}^n \ln (a_j) & \ell=0 \\
\f{2}{\ell} \lim_{m\to\infty} [\tr (T_\ell (\f12 J_{m;F}))-\tr(T_\ell (\f12 J_{m-n; F}^{(n)}))] 
& \ell=1,2,\dots 
\end{cases}
\end{equation}
where $J_{m;F}$ is the finite matrix formed from the first $m$ rows and columns of $J$ and 
$T_\ell$ is the $\ell$-th Chebyshev plynomial (of the first kind). As noted in 
\cite[Proposition~4.3]{KS}, the limit in \eqref{3.4} exists since the expression is independent   
of $m$ once $m>\ell +n$. 

Note that 
\begin{align}
\zeta_1^{(n)}(J) &= \sum_{j=1}^n b_j \lb{3.5} \\
\zeta_2^{(n)} (J) &= \sum_{j=1}^n \tfrac12 b_j^2 + (a_j^2 -1) \lb{3.6}
\end{align} 
as computed in \cite{KS}. 

Note that by construction (with $J^{(0)}\equiv J$), 
\begin{equation} \lb{3.7}
X_\ell^{(n)}(J) =\sum_{j=0}^{n-1} X_\ell^{(1)} (J^{(j)}) 
\end{equation}
and 
\begin{equation} \lb{3.8}
\zeta_\ell^{(n)}(J) = \sum_{j=0}^{n-1} \zeta_\ell^{(1)} (J^{(j)})
\end{equation}

As final objects we need 
\begin{equation} \lb{3.9}
Z(J) =\f{1}{4\pi}\, \int_0^{2\pi} \ln \biggl( \f{\sin\theta}{\Ima M(e^{i\theta}, J)}\biggr) 
\, d\theta
\end{equation}
and for $\ell \geq 1$, 
\begin{align} 
Z_\ell^\pm (J) &= \f{1}{4\pi} \, \int_0^{2\pi} \ln \biggl( \f{\sin\theta}{\Ima M(e^{i\theta}, J)} 
\biggr)\, (1\pm\cos (\ell\theta)) \, d\theta \lb{3.10} \\
Y_\ell (J) &= -\f{1}{2\pi}\, \int_0^{2\pi} \ln \biggl(\f{\sin\theta}{\Ima M(e^{i\theta},J)} 
\biggr)\, \cos (\ell\theta)\, d\theta \lb{3.11}
\end{align} 
We include ``$\sin\theta$" inside $\ln (\dots)$ so that $Z(J_0)=Z_\ell^\pm (J_0) =Y_\ell (J_0) =0$ 
because $M(z,J_0)=z$. 

Of course, 
\begin{equation} \lb{3.12}
Z_\ell^\pm (J) = Z(J) \mp \tfrac12\, Y_\ell (J)
\end{equation}
when all integrals converge. By Theorem~\ref{T2.2}, the $\ln_-$ piece of the integrals in 
\eqref{3.9}--\eqref{3.11} always converges. Since $1\pm\cos(\ell\theta)\geq 0$, the integrals 
defining $Z(J)$, $Z_\ell^\pm (J)$ either converge or diverge to $+\infty$. We therefore always 
define $Z(J)$ and $Z_\ell^\pm (J)$ although they may take the value $+\infty$. Since $[1\pm\cos 
(\ell\theta)]\leq 2$, $Z(J) <\infty$ implies $Z_\ell^\pm (J)<\infty$, so we define $Y_\ell 
(J)$ by \eqref{3.12} if and only if $Z(J)<\infty$. 

If $Z(J)<\infty$, we say $J$ obeys the Szeg\H{o} condition or $J$ is Szeg\H{o}. If $Z_1^\pm 
(J)<\infty$, we say $J$ is Szeg\H{o} at $\pm 2$ since, for example, if $Z_1^+ (J)<\infty$, the 
integral in \eqref{3.9} converges near $\theta =0$ ($E=2\cos(\theta)$ near $+2$) and if $Z_1^- 
(J)<\infty$, the integral converges near $\theta =\pi$ (i.e., $E=-2$). Note that while 
$Z_1^+ (J) <\infty$ only implies convergence of \eqref{3.9} at $\theta =0$, it also implies 
that at $\theta =\pi$ the integral with a $\sin^2 \theta$ inserted converges (quasi-Szeg\H{o} 
condition). 

Our main goal in this section is to prove the next three theorems 

\begin{theorem}[Step-by-Step Sum Rules] \lb{T3.1} Let $J$ be a BW matrix. $Z(J)<\infty$ if and 
only if $Z(J^{(1)})< \infty$, and if $Z(J)<\infty$, we have 
\begin{align} 
Z(J) &=-\ln (a_1) + X_0^{(1)} (J) + Z(J^{(1)}) \lb{3.13} \\
Y_\ell (J) &= \zeta_\ell^{(1)} (J) + X_\ell^{(1)}(J) + Y_\ell (J^{(1)}); 
\qquad \ell=1,2,3,\dots \lb{3.14} 
\end{align} 
\end{theorem}

{\it Remarks.} 1. By iteration and \eqref{3.7}/\eqref{3.8}, we obtain if $Z(J)<\infty$, then 
$Z(J^{(n)})<\infty$ and 
\begin{align}
Z(J) &= -\sum_{j=1}^n \ln (a_j) + X_0^{(n)}(J) + Z(J^{(n)}) \lb{3.15} \\
Y_\ell (J) &= \zeta_\ell^{(n)}(J) + X_\ell^{(n)}(J) + Y_\ell (J^{(n)}); 
\qquad \ell=1,2,3,\dots \lb{3.16}
\end{align}

\smallskip
2. We call \eqref{3.13}/\eqref{3.14} the step-by-step Case sum rules.

\begin{theorem}[One-Sided Step-by-Step Sum Rules]\lb{T3.2} Let $J$ be a BW matrix. $Z_1^\pm (J)
<\infty$ if and only if $Z_1^\pm (J^{(1)})<\infty$, and if $Z_1^\pm (J)<\infty$, then we have for 
$\ell=1,3,5, \dots$, 
\begin{equation} \lb{3.17}
Z_\ell^\pm (J) = -\ln (a_1)\mp \tfrac12\, \zeta_\ell^{(1)}(J) + X_0^{(1)} (J) \mp \tfrac12\, 
X_\ell^{(1)}(J) + Z_\ell^\pm (J^{(1)}) 
\end{equation}
\end{theorem}

{\it Remark.} Theorem~\ref{T3.2} is intended to be two statements: one with all the upper 
signs used and one with all the lower signs used.

\begin{theorem}[Quasi-Step-by-Step Sum Rules]\lb{T3.3} Let $J$ be a BW matrix. $Z_2^- (J)<
\infty$ if and only if $Z_2^- (J^{(1)})<\infty$, and if $Z_2^- (J)<\infty$, then for $\ell=
2,4,\dots$, we have 
\begin{equation} \lb{3.18}
Z_\ell^- (J) =-\ln (a_1) + \tfrac12\,\zeta_\ell^{(1)}(J) + X_0^{(1)}(J) + \tfrac12\, 
X_\ell^{(1)}(J) + Z_\ell^- (J^{(1)})
\end{equation}
\end{theorem}

{\it Remarks.} 1. The name comes from the fact that since $1-\cos 2\theta =2\sin^2\theta$, 
$Z_2^- (J)$ is what Killip-Simon \cite{KS} called the quasi-Szeg\H{o} integral
\begin{equation} \lb{3.19}
Z_2^- (J) =\f{1}{2\pi}\int_0^{2\pi} \ln \biggl(\f{\sin\theta}{\Ima M(e^{i\theta},J)}\biggr) 
\sin^2\theta\, d\theta
\end{equation}

\smallskip
2. Since $Z(J)<\infty$ implies $Z_1^+ (J)$ and $Z_1^- (J)<\infty$, and $Z_1^+(J)$ or $Z_1^-(J) 
<\infty$ imply $Z_2^- (J)<\infty$, we have additional sum rules in various cases.

\smallskip
3. In \cite{LNS}, Laptev et al.~prove sum rules for $Z_\ell^- (J)$ where $\ell=4,6,8,\dots$. 
One can develop step-by-step sum rules in this case and use it to streamline the proof of their 
rules as we streamline the proof of the Killip-Simon $P_2$ rule (our $Z_2^-$ sum rule) in the 
next section.

\smallskip
The step-by-step sum rules were introduced in Killip-Simon, who first take $r<1$ (in our 
language below), then take $n\to\infty$, and then $r\uparrow 1$ with some technical hurdles 
to take $r\uparrow 1$. By first letting $r\uparrow 1$ with $n<\infty$, and then $n\to\infty$ 
as in the next section, we can both simplify their proof and obtain additional results. The 
idea of using the imaginary part of 
\begin{equation} \lb{3.20}
-M(z;J)^{-1} = -(z+z^{-1}) + b_1 + a_1^2 M(z; J^{(1)})
\end{equation}
is taken from Killip-Simon \cite{KS}.

\begin{proof}[Proof of Theorem~\ref{T3.1}] Taking imaginary parts of both sides of \eqref{3.20} 
with $z=re^{i\theta}$ and $r<1$, we obtain 
\begin{equation} \lb{3.21}
[\Ima M(re^{i\theta};J)] \,\abs{M(re^{i\theta}; J)}^{-2} =(r^{-1}-r) \sin\theta + a_1^2 
\Ima M(re^{i\theta} ;J^{(1)})
\end{equation}
Taking $\ln$'s of both sides, we obtain 
\begin{equation} \lb{3.22}
\ln \biggl(\f{\sin\theta}{\Ima M(re^{i\theta};J)}\biggr) = t_1 + t_2 + t_3
\end{equation}
where 
\begin{align} 
t_1 &= -2\ln \abs{M(re^{i\theta};J)} \lb{3.23} \\
t_2 &= -2\ln a_1 \lb{3.24} \\
t_3 &= \ln \biggl( \f{\sin\theta}{g(r)\sin\theta + \Ima M(re^{i\theta};J^{(1)})}\biggr) \lb{3.25}
\end{align} 
where 
\begin{equation} \lb{3.26}
g(r) =a_1^{-2} (r^{-1}-r)
\end{equation}

Let 
\[
f(z) =\f{M(rz;J)}{rz}
\]
so $f(0)=1$ (see \eqref{3.20}). In the unit disk, $f(z)$ is meromorphic and has poles at 
$\{r(\beta_j^\pm (J))^{-1}\mid j\text{ so that } \beta_j^\pm (J) >r^{-1}\}$ and zeros at 
$\{r(\beta_j^\pm (J^{(1)}))^{-1} \mid \beta_j^\pm (J^{(1)})> r^{-1}\}$. Thus, by Jensen's 
formula for $f$: 
\[
\f{1}{4\pi}\int_0^{2\pi} t_1 \, d\theta =-\ln r + \sum_{\beta_j^\pm (J)>r^{-1}} \ln 
\abs{r^{-1} \beta_j^\pm (J)} -\sum_{\beta_j^\pm (J^{(1)}) >r^{-1}} \ln \abs{r^{-1} \beta_j^\pm 
(J^{(1)})}
\]
By \eqref{3.1}, the number of terms in the sums differs by at most $2$, so that the $\ln 
(r^{-1})$'s cancel up to at most $2\ln (r^{-1})\to 0$ as $r\uparrow 1$. Thus as $r\uparrow 1$, 
\begin{equation} \lb{3.27}
\f{1}{4\pi}\int_0^{2\pi} (t_1 + t_2) \, d\theta\to -\ln (a_1) + X_0^{(1)} (J)
\end{equation}

It follows by \eqref{3.22} and Theorems~\ref{T2.1} and \ref{T2.8} (with $w(\varphi)=1$) that 
$Z(J)<\infty$ if and only if $Z(J^{(1)})<\infty$, and if they are finite, \eqref{3.13} holds. 

It also follows that if $Z(J)<\infty$, we have $L^1$ convergence of the $\ln$'s to their 
$r=1$ values. That implies convergence of the integrals with $\cos (\ell\theta)$ inside. 
Higher Jensen's formula as in \cite{KS} then implies \eqref{3.14}. In place of $\ln\abs{\beta r^{-1}}$, 
we have $(r\beta)^\ell - (r\beta)^{-\ell}$, but the sums still converge to the $r=1$ limit 
since we can separate the $\beta^\ell$ and $\beta^{-\ell}$ terms, and then the $r$'s factor out. 
\end{proof} 

\begin{proof}[Proofs of Theorems~\ref{T3.2} and \ref{T3.3}] These are the same as the above proof, 
but now the weight $w$ is either $1\pm\cos(\theta)$ or $1-\cos(2\theta)$ and that weight obeys 
\eqref{2.3} and \eqref{2.4}. 
\end{proof}

\begin{corollary}\lb{C3.4} Let $J$ be a BW matrix. If $J$ and $\tilde J$ differ by a finite rank 
perturbation, then $J$ is Szeg\H{o} {\rm{(}}resp.~ Szeg\H{o} at $\pm 2${\rm{)}} if and only if 
$\tilde J$ is. 
\end{corollary}

\begin{proof} For some $n$, $J^{(n)}=\tilde J^{(n)}$, so this is immediate from 
Theorems~\ref{T3.1} and \ref{T3.2}. 
\end{proof}

\begin{conjecture}\lb{Con3.5} Let $J$ be a BW matrix. If $J$ and $\tilde J$ differ by a trace 
class perturbation, then $J$ is Szeg\H{o} (resp.~Szeg\H{o} at $\pm 2$) if and only if $\tilde J$ 
is. It is possible this conjecture is only generally true if $J-J_0$ is only assumed compact or 
is only assumed Hilbert-Schmidt.
\end{conjecture}

This conjecture for $J=J_0$ is Nevai's conjecture recently proven by Killip-Simon. Their 
method of proof and the ideas here would prove this conjecture if one can prove a result of 
the following form. Let $J,\tilde J$ differ by a finite rank operator so that by the 
discussion before \eqref{3.2}, 
\[
\lim_{N\to\infty} \sum_\pm \sum_{j=1}^N  \biggl( \sqrt{E_j^\pm (J)^2 -4\,} - 
\sqrt{E_j^\pm (\tilde J)^2-4\,}\,\biggr)  \equiv \delta (J, \tilde J)
\]
exists and is finite. The conjecture would be provable by the method of \cite{KS} and this paper 
(by using the step-by-step sum rule to remove the first $m$ pieces of $J$ and then replacing 
them with the first $m$ pieces of $\tilde J$) if one had a bound of the form 
\begin{equation} \lb{3.28}
\abs{\delta (J,\tilde J)} \leq \text{(const.)Tr} (\abs{J-\tilde J})
\end{equation}

\eqref{3.28} with $J=J_0$ is the estimate of Hundertmark-Simon \cite{HS}. We have counterexamples 
that show \eqref{3.28} does not hold for a universal constant $c$. However, in these examples, 
$\|J\|\to\infty$ as $c\to\infty$. Thus it could be that \eqref{3.28} holds with $c$ only depending on 
$J$ for some class of $J$'s. If it held with a bound depending only on $\|J\|$, the conjecture would 
hold in general. If $J$ was required in $J_0 +$ Hilbert-Schmidt, we would get the conjecture 
for such $J$'s. 

\bigskip
\section{The $Z_0$, $Z_1^\pm$, and $Z_2^-$ Sum Rules} \lb{s4}

Our goal here is to prove that sum rules of Case type hold under certain hypotheses. Of interest 
on their own, these considerations also somewhat simplify the proof of the $P_2$ sum rule in 
Section~8 of \cite{KS}, and considerably simplify the proof of the $C_0$ sum rule for trace class 
$J-J_0$ from Section~9 of \cite{KS}. Throughout, $J$ will be a BW matrix. There are two main tools. 
As in \cite{KS}, lower semicontinuity of the $Z$'s in $J$ (in the topology of pointwise convergence 
of matrix elements) gets inequalities in one direction. We use step-by-step sum rules and boundedness 
from below of $Z$ for the other direction. 

We first introduce some quantities involving a fixed Jacobi matrix:
\begin{align}
\bar A_0 (J) &=\limsup_{n\to\infty} \biggl(-\sum_{j=1}^n \ln (a_j)\biggr) \lb{4.1} \\
\ul{A}_0 (J)&= \liminf_{n\to\infty} \biggl(- \sum_{j=0}^n \ln (a_j)\biggr) \notag \\
\bar A_1^\pm (J) &= \limsup_{n\to\infty} \biggl(- \sum_{j=1}^n (a_j - 1\pm \tfrac12 b_j)\biggr) 
\lb{4.2} \\
\ul{A}_1^\pm (J) &= \liminf_{n\to\infty} \biggl(- \sum_{j=1}^n (a_j - 1 \pm\tfrac12 b_j)\biggr)
\notag \\
A_2 (J) &=\sum_{j=1}^\infty\, [\tfrac14 b_j^2 + \tfrac12 G(a_j)] \lb{4.3}
\end{align}
where 
\[
G(a) =a^2 - 1- \ln (a^2)
\]
Since $G(a)\geq 0$, the finite sums have a limit (which may be $+\infty$). 

We note that for $a$ near $1$, $G(a)\sim 2(a-1)^2$. Thus $A_2 (J)$ is finite if and only if $J-J_0$ 
is Hilbert-Schmidt. In \eqref{4.2}, we can use $a_j -1$ in place of $\ln (a_j)$ because if $\{a_j 
-1\} \in \ell^2$ (e.g., if $J-J_0$ is Hilbert-Schmidt), then $\sum \abs{\ln (a_j) - (a_j -1)} 
<\infty$.

We also introduce some functions of the eigenvalues: 
\begin{align}
\calE_0 (J) &= \sum_{j,\pm} \ln \abs{\beta_j^\pm} \lb{4.4} \\
\calE_1^\pm (J) &= \sum_j \sqrt{(E_j^\pm)^2 -4\,} \lb{4.5} \\
\calE_2 (J) &= \sum_{j,\pm} F(E_j^\pm) \lb{4.6} 
\end{align}
where $F(E)=\f14 [\beta^2 - \beta^{-2}-\ln (\beta^4)]$ with $E=\beta + \beta^{-1}$ and 
$\abs{\beta}>1$. For $\abs{E}\sim 2$, $F(E)$ is $O((\abs{E}-2)^{3/2})$. In \eqref{4.4} and 
\eqref{4.6}, we sum over $+$ and $-$. In \eqref{4.5}, we define $\calE_1^+$ and $\calE_1^-$ with 
only the $+$ or only the $-$ terms. 

We need the following basis-dependent notion:

\smallskip
\noindent{\bf Definition.} Let $B$ be a bounded operator on $\ell^2 (\bbZ^+)$. We say $B$ has 
a conditional trace if 
\begin{equation} \lb{4.6a}
\lim_{\ell\to\infty} \, \sum_{j=1}^\ell \langle\delta_j, B\delta_j\rangle \equiv \text{c-Tr}(B)
\end{equation}
exists and is finite.

If $B$ is not trace class, this object is not unitarily invariant.

Our goal in this section is to prove the following theorems whose proof is deferred until after 
all the statements.

\begin{theorem}\lb{T4.1} Let $J$ be a BW matrix. Consider the four statements:
\begin{SL}
\item[{\rm{(i)}}] $\bar A_0 (J) > -\infty$
\item[{\rm{(ii)}}] $\ul{A}_0 (J) <\infty$ 
\item[{\rm{(iii)}}] $Z(J) <\infty$ 
\item[{\rm{(iv)}}] $\calE_0 (J) <\infty$
\end{SL}
Then 
\begin{SL} 
\item[{\rm{(a)}}] $\text{\rm{(ii)}} + \text{\rm{(iv)}} \Rightarrow \text{\rm{(iii)}} + 
\text{\rm{(i)}}$
\item[{\rm{(b)}}] $\text{\rm{(i)}} + \text{\rm{(iii)}} \Rightarrow \text{\rm{(iv)}} + 
\text{\rm{(ii)}}$
\item[{\rm{(c)}}] $\text{\rm{(iii)}} \Rightarrow \bar A_0 (J) <\infty$
\item[{\rm{(d)}}] $\text{\rm{(iv)}} \Rightarrow \ul{A}_0 (J) > -\infty$
\end{SL}
Thus $\text{\rm{(iii)}} + \text{\rm{(iv)}} \Rightarrow \text{\rm{(i)}} + \text{\rm{(ii)}}$. 
In particular, if $\ul{A}_0 (J) =\bar A_0 (J)$, that is, the limit exists, then the finiteness 
of any two of $Z(J)$, $\calE_0(J)$, and $\bar A_0(J)$ implies the finiteness of the third.

If all four conditions hold and $J-J_0$ is compact, then 
\begin{SL}
\item[{\rm{(e)}}] 
\begin{equation} \lb{4.7}
\lim_{n\to\infty}\, \biggl( -\sum_{j=1}^n \, \ln(a_j)\biggr) \equiv A_0(J)
\end{equation}
exists and is finite, and 
\begin{equation} \lb{4.8}
Z(J) =A_0(J) +\calE_0 (J)
\end{equation}
\item[{\rm{(f)}}] For each $\ell=1,2,\dots$, 
\begin{equation} \lb{4.9}
-\sum_{j,\pm} \ell^{-1} [\beta_j^\pm (J)^\ell - \beta_j^\pm (J)^{-\ell}] \equiv 
X_\ell^{(\infty)}(J)
\end{equation}
converges absolutely and equals $\lim_{n\to\infty} X_\ell^{(n)} (J)$. 
\item[{\rm{(g)}}] For each $\ell=1,2,\dots$,   
\begin{equation} \lb{4.10}
B_\ell (J) =\f{2}{\ell} \biggl\{T_\ell \biggl(\f{J}{2}\biggr) - T_\ell \biggl(\f{J_0}{2}\biggr) 
\biggr\}
\end{equation}
has a conditional trace and 
\begin{equation} \lb{4.11}
\text{\rm{c-Tr}} (B_\ell (J))=\lim_{n\to\infty}\, \zeta_\ell^{(n)}(J) 
\end{equation}
for example, if $\ell=1$, $\sum_{j=1}^n b_j$ converges to a finite limit. 
\item[{\rm{(h)}}] The Case sum rule holds: 
\begin{equation} \lb{4.12}
Y_\ell (J) = \text{\rm{c-Tr}} (B_\ell (J)) + X_\ell^{(\infty)}(J)
\end{equation}
where $Y_\ell$ is given by \eqref{3.11}, $X_\ell^{(\infty)}$ by \eqref{4.9}, and $\text{\rm{c-Tr}}
(B_\ell (J))$ by \eqref{4.6a}, \eqref{4.10}, and \eqref{4.11}.
\end{SL}
\end{theorem} 

{\it Remarks.} 1. In one sense, this is the main result of this paper. 

\smallskip
2. We will give examples later where $\bar A_0 (J)=\ul{A}_0 (J)$ and one of the conditions (i)/(ii), 
(iii), (iv) holds and the other two fail.

\smallskip
3. For $\ell$ odd, $T_\ell (J_0/2)$ vanishes on-diagonal. By Proposition~2.2 of \cite{KS} 
and the fact that the diagonal matrix elements of $J_0^k$ are eventually constant, it follows 
that for $\ell$ even, $T_\ell (J_0/2)$ eventually vanishes on-diagonal and $\text{c-Tr} (T_\ell 
(J_0/2))=-\f12$. Thus (g) says $\text{c-Tr} (T_\ell (J/2))$ exists and the sum rule \eqref{4.12} 
can replace $\text{c-Tr} (B_\ell (J))$ by $\text{c-Tr}(T_\ell (J/2))$ plus a constant (zero if 
$\ell$ is odd and $1/\ell$ if $\ell$ is even). For $\ell$ even, $\text{c-Tr}(T_\ell (J_0/2))=-\f12$ 
while $\text{Tr} (T_\ell (J_{0,n;F}/2))=-1$ for $n$ large because $T_\ell (J_{0,n;F}/2)$ 
has two ends. 

\begin{corollary} \lb{C4.2} Let $J-J_0$ be compact. If $Z(J)<\infty$, then $-\sum_{j=1}^n  
\ln (a_j)$ either converges or diverges to $-\infty$. 
\end{corollary} 

{\it Remarks.} 1. We will give an example later where $Z(J)<\infty$, and $\lim_{n\to\infty} 
(-\sum_{j=1}^n \ln (a_j))=-\infty$. 

\smallskip
2. In other words, if $J-J_0$ is compact and $\bar A_0 (J)\neq \ul{A}_0 (J)$, then $Z(J)=\infty$. 

\smallskip
3. Similarly, if $J-J_0$ is compact and $\calE_0 (J)<\infty$, then the limit exists and is 
finite or is $+\infty$. 

\begin{proof} If $Z(J)<\infty$ and $\bar A_0 >-\infty$, then by (b) of the theorem, all four 
conditions hold, and so by (e), the limit exists. On the other hand, if $\bar A_0 =-\infty$, 
then $\bar A_0=\ul{A}_0 =-\infty$. 
\end{proof}

\begin{corollary} \lb{C4.3} If $J-J_0$ is trace class, then $Z(J)<\infty$, $\calE_0 (J)<\infty$, 
and the sum rules \eqref{4.8} and \eqref{4.12} hold. 
\end{corollary} 

{\it Remark.} This is a result of Killip-Simon \cite{KS}. Our proof that $Z(J)<\infty$ is 
essentially the same as theirs, but our proof of the sum rules is much easier.

\begin{proof} Since $J-J_0$ is trace class, it is compact. Clearly, $\bar A_0 =\ul{A}_0$, 
and is neither $\infty$ nor $-\infty$ since $a_j >0$ and $\sum\abs{a_j- 1}<\infty$ imply  
$\sum\abs{\ln (a_j)}<\infty$. By the bound of Hundertmark-Simon \cite{HS}, $\calE_0 (J) 
<\infty$. The sum rules then hold by (a), (e), and (h) of Theorem~\ref{T4.1}. 
\end{proof}

\begin{theorem}\lb{T4.4} Suppose $J-J_0$ is Hilbert-Schmidt. Then 
\begin{SL} 
\item[{\rm{(i)}}] $\ul{A}_1^\pm <\infty$ and $\calE_1^\pm <\infty$ implies $Z_1^\pm <\infty$. 
\item[{\rm{(ii)}}] $Z_1^\pm <\infty$ implies $\bar A_1^\pm <\infty$.
\item[{\rm{(iii)}}] $Z_1^\pm <\infty$ and $\bar A_1^\pm >-\infty$ implies $\calE_1^\pm <\infty$. 
\item[{\rm{(iv)}}] $\calE_1^\pm <\infty$ implies $\ul{A}_1^\pm >-\infty$.
\end{SL}
\end{theorem}

{\it Remarks.} 1. Each of (i)--(iv) is intended as two statements. 

\smallskip
2. In Section~\ref{s6}, we will explore (ii), which is the most striking of these 
results since its contrapositive gives very general conditions under which the Szeg\H{o} condition 
fails.

\smallskip
3. The Hilbert-Schmidt condition in (i) and (iv) can be replaced by the somewhat weaker condition that 
\begin{equation} \lb{4.13}
\sum_{j,\pm} \, (\abs{E_j^\pm}-2)^{3/2} <\infty 
\end{equation}
That is true for (ii) and (iii) also, but by the $Z_2^-$ sum rule, \eqref{4.13} plus $Z_1^\pm < 
\infty$ implies $J-J_0$ is Hilbert-Schmidt.

\begin{theorem} \lb{T4.5} Let $J$ be a BW matrix. Then  
\begin{equation} \lb{4.13x}
Z_2^- (J) + \calE_2 (J) = A_2 (J) 
\end{equation}
\end{theorem}

{\it Remarks.} 1. This is, of course, the $P_2$ sum rule of Killip-Simon \cite{KS}. Our proof that 
$Z_2 ^- (J) + \calE_2 (J)\leq A_2 (J)$ is identical to that in \cite{KS}, but our proof of the other 
half is somewhat streamlined.

\smallskip
2. As in \cite{KS}, the values $+\infty$ are allowed in \eqref{4.13x}. 

\begin{proof}[Proof of Theorem~\ref{T4.1}] As in \cite{KS}, let $J_n$ be the infinite Jacobi matrix 
obtained from $J$ by replacing $a_\ell$ by $1$ if $\ell\geq n$ and $b_\ell$ by $0$ if $\ell \geq 
n+1$. Then \eqref{3.15} (noting $J_n^{(n)}=J_0$ and $Z(J_0)=0$) reads
\begin{equation} \lb{4.14}
Z(J_n) =-\sum_{j=1}^n \ln (a_j) + \sum_{j,\pm} \ln\abs{\beta_j^\pm (J_n)}
\end{equation}

\cite[Section 6]{KS} implies the eigenvalue sum converges to $\calE_0 (J)$ if $J-J_0$ is compact, 
and in any event, is bounded above by $\calE_0 (J) +c_0$ where $c_0=0$ if $J-J_0$ is compact and 
otherwise, 
\begin{equation} \lb{4.15a}
c_0  =\ln \abs{\beta_1^+ (J)} + \ln\abs{\beta_1^-(J)}
\end{equation}
Moreover, by semicontinuity of the entropy \cite[Section 5]{KS}, $Z(J)\leq\liminf Z(J_n)$. Thus 
we have 
\begin{equation} \lb{4.15}
Z(J) \leq \ul{A}_0(J)+\calE_0(J)+c_0
\end{equation}

Thus far, the proof is directly from \cite{KS}. On the other hand, by \eqref{3.15}, we have 
\begin{equation} \lb{4.16}
Z(J) \geq \bar A_0 (J) + \liminf X_0^{(n)}(J) + \liminf Z(J^{(n)})
\end{equation}
By the lemma below, $\lim_{n\to\infty} X_0^{(n)}(J)=\calE_0(J)$. Moreover, by Theorem~5.5 
(eqn.~(5.26)) of Killip-Simon \cite{KS}, $Z(J^{(n)})\geq -\f12 \ln(2)$, and if $J^{(n)}\to 
J_0$ in norm, that is, $J-J_0$ is compact, then by semicontinuity of $Z$, $0=Z(J_0)\leq\liminf 
Z(J^{(n)})$. Therefore, \eqref{4.16} implies that 
\begin{equation} \lb{4.17}
Z(J) \geq \bar A_0 (J)+\calE_0 (J)-c
\end{equation}
where 
\begin{equation} \lb{4.18}
c=0 \text{ if $J-J_0$ is compact}; \qquad c=\tfrac12 \ln 2 \text{ in general}
\end{equation}

With these preliminaries out of the way, 

\smallskip
\noindent\ul{Proof of (d).} \ (iv) and \eqref{4.15} imply that 
\begin{equation} \lb{4.18a}
\bar A_0(J) \geq \ul{A}_0 (J) \geq Z(J) -\calE_0 (J) -c_0 >-\infty
\end{equation}

\smallskip
\noindent\ul{Proof of (a).} \ \eqref{4.15} shows $Z(J) <\infty$, and (d) shows that (i) holds.

\smallskip
\noindent\ul{Proof of (c).} \ By \eqref{4.17} and $\calE_0 (J) \geq 0$, 
\[
Z(J) \geq \bar A_0 (J) -c 
\]
so $Z(J)<\infty$ implies $\bar A_0 (J)<\infty$. 

\smallskip
\noindent\ul{Proof of (b).} \ Since $\bar A_0(J) >-\infty$ and $c<\infty$, \eqref{4.17} plus $Z(J) 
<\infty$ implies $\calE_0 (J)<\infty$. (c) shows that (ii) holds. 

\smallskip
Note that (iii), (iv), and \eqref{4.17} imply that 
\begin{equation} \lb{4.19}
\ul{A}_0(J) \leq \bar A_0 (J) \leq Z(J) -\calE_0 (J) + \tfrac12 \ln 2 <\infty
\end{equation}
Thus we have shown more than merely (iii) + (iv) $\Rightarrow$ (i) + (ii), namely, (iii) + 
(iv) imply by \eqref{4.18a} and \eqref{4.19} 
\begin{equation} \lb{4.20}
-\infty < \bar A_0(J) \leq \ul{A}_0 (J) + \tfrac12 \ln 2 + c_0<\infty
\end{equation}
We can say more if $J-J_0$ is compact. 

\smallskip
\noindent\ul{Proof of (e).} \ \eqref{4.19} is now replaced by 
\[
\ul{A}_0 (J) \leq \bar A_0 (J) \leq Z(J) - \calE_0 (J)
\]
since we can take $c=0$ in \eqref{4.17}. This plus \eqref{4.18a} with $c_0=0$ implies 
$\bar A_0 (J) = \ul{A}_0 (J)$ and \eqref{4.8}.

\smallskip
\noindent\ul{Proof of (f), (g), (h).} \ We have the sum rules \eqref{3.15}, \eqref{3.16}. $Z(J) 
\pm Y_\ell (J)$ is an entropy up to a constant, and so, lower semicontinous. Since 
$\|J^{(n)}-J_0\| \to 0$, we have 
\begin{equation} \lb{4.21}
\liminf (Z(J^{(n)}) \pm Y_\ell (J^{(n)})) \geq 0
\end{equation}
On the other hand, since $Z(J^{(n)})<\infty$ and $\calE_0 (J^{(n)}) \leq \calE_0 (J) <\infty$, 
$J^{(n)}$ obeys the sum rule \eqref{4.8}. Since $-\sum_{j=1}^n \ln (a_j)$ converges conditionally 
\[
\lim_{n\to\infty} \lim_{m\to\infty} \biggl(-\sum_{j=n}^{m+n} \ln (a_j)\biggr) =0
\]
Moreover, $\calE_0 (J^{(n)})\to 0$ by Lemma~\ref{L4.6} below and we conclude that $\lim Z(J^{(n)})
=0$. Thus \eqref{4.21} becomes 
\[
\liminf_n \, Y_\ell (J^{(n)})\geq 0, \qquad \limsup_n \, Y_\ell (J^{(n)})\leq 0
\]
or
\begin{equation} \lb{4.22}
\lim_n\, Y_\ell (J^{(n)})=0
\end{equation}

By the lemma below, $\lim_n X_\ell^{(n)}(J) =X_\ell^{(\infty)}(J)$ exists and is finite. Since 
$\calE_0 (J)<\infty$, we have that the sum defining $X_\ell^{(\infty)}(J)$ is absolutely 
convergent. This proves (f).

By this fact, \eqref{3.16}, and \eqref{4.22}, $\lim_{n\to\infty} \zeta_\ell^{(n)} (J)$ exists,  
is finite, and obeys the sum rule 
\[
Y_\ell (J) =\lim_{n\to\infty} \, \zeta_\ell^{(n)}(J) + X_\ell^{(\infty)}(J)
\]

By Propositions~2.2 and 4.3 of Killip-Simon \cite{KS}, the existence of $\lim_{n\to\infty} 
\zeta_\ell^{(n)} (J)$ is precisely the existence of the conditional trace. 
\end{proof}

\begin{lemma}\lb{L4.6} Let $J$ be a BW matrix. Let $f$ be a monotone increasing continuous function  
on $[2,\infty)$ with $f(2)=0$. Then 
\begin{equation} \lb{4.23}
\lim_{n\to\infty} \,  \sum_{j=1}^\infty \, [f(E_j^+ (J)) - f(E_j^+(J^{(n)}))] = \sum_{j=1}^\infty 
f(E_j^+ (J))
\end{equation}
\end{lemma}

{\it Remarks.} 1. The right side of \eqref{4.23} may be finite or infinite.

\smallskip
2. The sum on the left is interpreted as the limit of the sum from $1$ to $n$ as $n\to\infty$, 
which exists and is finite by the arguments at the start of Section 3.

\smallskip
3. A similar result holds for $E_j^-$ and $f$ monotone decreasing on $(-\infty, -2]$.

\begin{proof} Call the sum on the left of \eqref{4.23} $(\delta f)(J,n)$. Since $E_j^+ (J^{(n)}) 
\leq E_j^+ (J)$, we have 
\begin{equation} \lb{4.24}
(\delta f)(J,n) \geq \sum_{j=1}^m \, [f(E_j^+ (J))-f(E_j^+ (J^{(n)}))]
\end{equation}
so, if we show for each fixed $j$ as $n\to\infty$, 
\begin{equation} \lb{4.25}
E_j^+ (J^{(n)})\to 2
\end{equation}
we have, by taking $n\to\infty$ and then $m\to\infty$, that 
\begin{equation} \lb{4.26}
\liminf (\delta f)(J,n) \geq \sum_{j=1}^\infty f(E_j^+ (J))
\end{equation}

On the other hand, since $f\geq 0$, for each $m$, 
\[
\sum_{j=1}^m \, [f(E_j^+ (J)) - f(E_j^+ (J^{(n)}))] \leq \sum_{j=1}^m f(E_j^+ (J))
\]
so taking $m$ to infinity and then $n\to\infty$, 
\begin{equation} \lb{4.27}
\limsup (\delta f)(J,n) \leq \sum_{j=1}^\infty f(E_j^+ (J))
\end{equation}
Thus \eqref{4.25} implies the result, so we need only prove that.

Fix $\veps >0$ and look at the solution of the orthogonal polynomial sequence $u_n =P_n 
(2+\veps)$ as a function of $n$. By Sturm oscillation theory \cite{FP}, the number of sign 
changes of $u_n$ (i.e., number of zeros of the piecewise linear interpolation of $u_n$) is 
the number of $j$ with $E_j^+ (J) >2+\veps$. Since $J$ is a BW matrix, this is finite, so 
there exist $N_0$ with $u_n$ of definite sign if $n\geq N_0 -1$. It follows by Sturm 
oscillation theory again that for all $j$, 
\[
E_j^+ (J^{(n)})\leq 2+\veps
\]
if $n\geq N_0$. This implies \eqref{4.25}. 
\end{proof}

The combination of this Sturm oscillation argument and Theorem~\ref{T3.1} gives one tools to 
handle finitely many bound states as an alternate to Nikishin \cite{Nik}. For the oscillation 
argument says that if $J$ has finitely many eigenvalues outside $[-2,2]$, there is a $J^{(n)}$ 
with no eigenvalues. On the other hand, by Theorem~\ref{T3.1}, $Z(J) <\infty$ if and only 
if $Z(J^{(n)})<\infty$. 

\begin{proof}[Proof of Theorem~\ref{T4.5}] $Z_2^- (J)$ is an entropy and not merely an entropy 
up to a constant (see \cite{KS}). Thus $Z_2^- (J^{(n)})\geq 0$ for all $J^{(n)}$. Moreover, since 
the terms in $A_2$ are positive, the limit exists. Thus, following the proofs of \eqref{4.15} 
and \eqref{4.17} but using \eqref{3.18} in place of \eqref{3.15}, 
\[
Z_2^- (J) + \calE_2 (J) \leq A_2 (J)
\]
and 
\[
Z_2^- (J) + \calE_2 (J) \geq A_2 (J)
\]
which yields the $P_2$ sum rule. In the above, we use the fact that in place of $Z(J)\geq -\f12 
\ln (2)$, one has $Z_2^- (J) \geq 0$, and the fact that $A_2 (J)<\infty$ implies that $J-J_0$ 
is compact. 
\end{proof}

\begin{proof}[Proof of Theorem~\ref{T4.4}] Let $g(\beta)=\ln\beta-\f12 (\beta-\beta^{-1})$ in 
 the region $\beta >0$. Then 
\[
g'(\beta) =\beta^{-1} -\tfrac12 - \tfrac12 \beta^{-2} = -\tfrac12\beta^{-2} (\beta -1)^2 
\]
so $g$ is analytic near $\beta =1$ and $g(1)=g'(1)=g''(1)=0$, that is, $g(\beta) \sim c
(\beta -1)^3$. On the other hand, $h(\beta) =\ln \beta + \tfrac 12 (\beta - \beta^{-1})$ 
is $g(\beta) + (\beta - \beta^{-1})=\beta - \beta^{-1} +O((\beta-1)^3)$. Since $\beta + \beta^{-1} 
=E$ means $\beta - \beta^{-1}=\sqrt{E^2 -4}$ and $\beta -1 =O\bigl(\sqrt{E-2\,}\bigr)$, 
we conclude that 
\begin{align*}
E>2 \Rightarrow &\ln(\beta)-\tfrac12 (\beta - \beta^{-1}) = O(\abs{E-2}^{3/2}) \\
&\ln (\beta)+\tfrac12 (\beta - \beta^{-1}) = \sqrt{E^2 -4} +O(\abs{E-2}^{3/2})
\end{align*}
while 
\begin{align*}
E< -2 \Rightarrow &\ln (\abs{\beta}) -\tfrac12 (\beta - \beta^{-1}) =\sqrt{E^2 -4} + 
O(\abs{E+2}^{3/2}) \\
&\ln (\abs{\beta}) + \tfrac12 (\beta - \beta^{-1}) =O(\abs{E+2}^{3/2}) 
\end{align*}

It follows, using Lemma~\ref{L4.6}, that 
\[
\lim_{n\to\infty}\, X_0^{(n)}(J)\mp \tfrac12 X_1^{(n)}(J) = \calE_1^\pm + \text{bdd} 
\]
since Theorem~\ref{T4.5} implies $\sum_{j,\pm} \bigl(\sqrt{E^{\pm 2}_j -4\,}\bigr)^3<\infty$ 
(or, by results of \cite{HS}). Thus for a constant $c_1$ dependng only on $\|J-J_0\|_2$, we have 
\begin{equation} \lb{4.28}
Z_1^\pm (J) \leq c_1 + \ul{A}_1^\pm + \calE_1^\pm
\end{equation}
by writing the finite rank sum rule, taking limits and using the argument between \eqref{4.14} 
and \eqref{4.15a}. Since $Z_1^\pm (J)$ are entropies up to a constant, we have $Z_1^\pm (J^{(n)})
\geq -c_2$ and so by \eqref{3.17}, 
\begin{equation} \lb{4.29}
Z_1^\pm (J) \geq -c_2 + \bar A_1^\pm + \calE_1^\pm -c\|J-J_0\|_2^2
\end{equation}

With these preliminaries, we have 

\smallskip
\noindent\ul{Proof of (i), (iv).} \ Immediate from \eqref{4.28}. 

\smallskip
\noindent\ul{Proof of (ii).} \ Since $\calE_1^\pm \geq 0$, \eqref{4.29} implies 
\[
Z_1^\pm (J) \geq -c_2 + \bar A_1^\pm
\]
so (ii) holds.

\smallskip
\noindent\ul{Proof of (iii).} \ Immediate from \eqref{4.29}. 
\end{proof}

\smallskip
{\it Remark.} (i)--(iv) of Theorem~\ref{T4.4} are exactly (a)--(d) of Theorem~\ref{T4.1} for 
the $Z_1^{\pm}$ sum rules. One therefore expects a version of (e) of that theorem to hold as well. 
Indeed, a modification of the above proof yields for $J-J_0$ Hilbert-Schmidt that if $\calE_1^{+}$, 
$Z_1^+$, $\bar A_1^+$ are finite, then
\[
Z_1^+(J)=-\sum_{n=1}^\infty \, [\ln(a_n)+\tfrac12 b_n]+\sum_{j,\pm} \, [\ln\abs{\beta_j^\pm}
+\tfrac12(\beta_j^\pm-(\beta_j^\pm)^{-1})]
\]
and if $\calE_1^{-}$, $Z_1^-$, $\bar A_1^-$ are finite, then
\[
Z_1^-(J)=-\sum_{n=1}^\infty \, [\ln(a_n)-\tfrac12 b_n]+\sum_{j,\pm} \, [\ln\abs{\beta_j^\pm}
-\tfrac12(\beta_j^\pm-(\beta_j^\pm)^{-1})]
\]

\bigskip
\section{Shohat's Theorem with an Eigenvalue Estimate} \lb{s5}

Shohat \cite{Sh} translated Szeg\H{o}'s theory from the unit circle to the real line and was able 
to identify all Jacobi matrices which lead to measures with no mass points outside $[-2,2]$ and 
have $Z(J)<\infty$. The strongest result we know of this type is the following (Theorem~$4'$) 
from Killip-Simon \cite{KS} (the methods of Nevai \cite{NevAMS} can prove the same result): 

\begin{theorem} \lb{T5.1} Let $\sigma(J)\subset [-2,2]$. Consider 
\begin{SL} 
\item[{\rm{(i)}}] $\ul{A}_0 (J) < \infty$ where $\ul{A}_0$ is given by \eqref{4.1}. 
\item[{\rm{(ii)}}] $Z(J) <\infty$ 
\item[{\rm{(iii)}}] $\sum_{n=1}^\infty (a_n -1)^2 + \sum_{n=1}^\infty b_n^2 <\infty$ 
\item[{\rm{(iv)}}] $\ul{A}_0 = \bar A_0$ and is finite. 
\item[{\rm{(v)}}] $\lim_{N\to\infty} \sum_{n=1}^N b_n$ exists and is finite. 
\end{SL}
Then {\rm{(}}under $\sigma(J) \subset [-2,2]${\rm{)}}, we have 
\[
\text{\rm{(i)}} \Longleftrightarrow \text{\rm{(ii)}}
\]
and either one implies {\rm{(iii)}}, {\rm{(iv)}}, and {\rm{(v)}}.
\end{theorem}

We can prove the following extension of this result:

\begin{theorem} \lb{T5.2} Theorem~\ref{T5.1} remains true if $\sigma (J) \subset [-2,2]$ is 
replaced by $\sigma_{\ess} (J) \subset [-2,2]$ and \eqref{1.5new}. 
\end{theorem}

{\it Remarks.} 1. Gon\v{c}ar \cite{Gon}, Nevai \cite{NevAMS}, and Nikishin \cite{Nik} extended 
Shohat-type theorems to allow finitely many bound states outside $[-2,2]$. 

\smallskip
2. Peherstorfer-Yuditskii \cite{PY} recently proved that $\calE_0(J)<\infty$ and (ii) implies (iv) 
and additional results on polynomial asymptotics. 

\begin{proof} Let us suppose first $\sigma_{\ess}(J)=[-2,2]$, so $J$ is a BW matrix. By 
Theorem~\ref{T4.1}(a), (i) of this theorem plus $\calE_0 (J)<\infty$ implies (ii) of this 
theorem. By Theorem~\ref{T4.1}(c), (ii) of this theorem implies (i) of this theorem.

If either holds, then (iv) follows from (e) of Theorem~\ref{T4.1}, (v) from the $\ell=1$ case 
of (g) of Theorem~\ref{T4.1}. (iii) follows from Theorem~\ref{T4.5} if we note that $\calE_0
<\infty$ implies $\calE_2 <\infty$, that $Z(J)<\infty$ implies $Z_2^-(J)<\infty$ and that 
$G(a)=O((a-1)^2)$. 

If we only have a priori that $\sigma_{\ess}(J)\subset [-2,2]$, we proceed as follows. If $Z(J) 
<\infty$, $\sigma_{\ac}(J)\supset [-2,2]$ so, in fact, $\sigma_{\ess}(J)=[-2,2]$. If $\ul{A}_0 
<\infty$, we look closely at the proof of Theorem~\ref{T4.1}(a). \eqref{4.15} does not 
require $\sigma_{\ess}(J)=[-2,2]$, but only that $\sigma_{\ess}(J) \subset [-2,2]$. Thus, 
$\ul{A}_0 <\infty$ implies $Z(J)<\infty$ if $\calE_0 (J)<\infty$. 
\end{proof}

There is an interesting way of rephrasing this. Let the normalized orthogonal polynomial obey 
\begin{equation} \lb{5.1}
P_n (x) = \gamma_n x^n +O(x^{n-1})
\end{equation}
As is well known (see, e.g. \cite{SimM}), 
\begin{equation} \lb{5.2}
\gamma_n =(a_1 a_2 \dots a_n)^{-1}
\end{equation}
Thus
\begin{equation} \lb{5.3}
\ul{A}_0 =\liminf \ln (\gamma_n)
\end{equation}
and 
\begin{equation} \lb{5.4}
\bar A_0 =\limsup \ln (\gamma_n)
\end{equation}

\begin{corollary} \lb{C5.2} Suppose $\sigma_{\ess}(J) \subset [-2,2]$ and $\calE_0 (J)<\infty$. 
Then $Z(J)<\infty$ {\rm{(}}i.e., the Szeg\H{o} condition holds{\rm{)}} if and only if $\gamma_n$ 
is bounded from above {\rm{(}}and in that case, it is also bounded away from $0$; indeed, 
$\lim \gamma_n$ exists and is in $(0,\infty)${\rm{)}}.
\end{corollary} 

{\it Remark.} Actually, $\limsup \gamma_n<\infty$ is not needed; $\liminf \gamma_n <\infty$ is enough. 

\begin{proof} By \eqref{5.3}, $\gamma_n$ bounded implies $\ul{A}_0 <\infty$, and thus $Z(J)<\infty$. 
Conversely, $Z(J)<\infty$ implies $-\infty <\ul{A}_0 = \bar A_0 <\infty$. So by \eqref{5.2}, it 
implies $\gamma_n$ is bounded above and below. 
\end{proof}

In the case of orthogonal polynomials on the circle, Szeg\H{o}'s theorem says $Z<\infty$ if and 
only if $\kappa_j$ is bounded if and only if $\sum_{j=1}^\infty \abs{\alpha_j}^2 <\infty$ where 
$\kappa_j$ is the leading coefficient of the normalized polynomials, and $\alpha_j$ are the 
Verblunsky (aka Geronimus, aka reflection) coefficients. In the real line case, if one drops 
the a priori requirement that $\calE_0 (J)<\infty$, it can happen that $\gamma_n$ is bounded but 
$Z(J) =\infty$. For example, if $a_n \equiv 1$ but $b_n =n^{-1}$, then $Z(J)$ cannot be finite. 
For $J-J_0 \in\ell_2$, so Theorem~\ref{T4.4}(ii) is applicable and thus, $\bar A_1^- =\infty$ 
implies $Z(J)=\infty$. 

But the other direction always holds:

\begin{theorem} \lb{T5.3} Let $J$ be a BW matrix with $Z(J)<\infty$ {\rm{(}}i.e., the Szeg\H{o} 
condition holds{\rm{)}}. Then $\gamma_n$ is bounded. Moreover, if $J-J_0$ is compact, then 
$\lim_{n\to\infty} \gamma_n$ exists.
\end{theorem}

{\it Remarks.} 1. The examples of the next section show $Z(J)<\infty$ is consistent with $\lim 
\gamma_n =0$. 

\smallskip
2. This result --- even without a compactness hypothesis --- is known. For $\gamma_n$ is 
monotone increasing in the measure (see, e.g., Nevai \cite{NF}) and so one can reduce to the case 
where Shohat's theorem applies. 

\begin{proof} By Theorem~\ref{T4.1}(c), $Z(J)<\infty$ implies $\bar A_0 <\infty$ which, by 
\eqref{5.4}, implies $\gamma_n$ is bounded. If $J-J_0$ is compact, then Corollary~\ref{C4.2} 
implies that $\lim \gamma_n =\exp (\lim -\sum_{j=1}^m \ln (a_j))$ exists but can be zero. 
\end{proof}

Here is another interesting application of Theorem~\ref{T5.2}.

\begin{theorem}\lb{T5.5} Suppose $b_n\geq 0$ and 
\begin{equation} \lb{5.5}
\sum_{n=1}^\infty \, \abs{a_n-1} <\infty
\end{equation}
Then $\calE_0 (J)<\infty$ if and only if $\sum_{n=1}^\infty b_n<\infty$. 
\end{theorem}

\begin{proof} If $\sum_{n=1}^\infty b_n <\infty$, $\calE_0 (J)<\infty$ by \eqref{5.5} and the 
bounds of Hundertmark-Simon \cite{HS}. On the other hand, if $\calE_0 (J)<\infty$, \eqref{5.5} implies 
$\ul{A}_0<\infty$, so by Theorem~\ref{T5.2}, $\sum_{n=1}^N b_n$ is convergent. Since $b_n \geq 0$, 
$\sum_{n=1}^\infty b_n <\infty$. 
\end{proof}

\bigskip
\section{$O(n^{-1})$ Perturbations} \lb{s6}

In this section, we will discuss examples where 
\begin{align}
a_n &= 1 + \alpha n^{-1} + E_a (n) \lb{6.1} \\
b_n &= \beta n^{-1} + E_b (n) \lb{6.2}
\end{align}
where $E_\cdot (n)$ is small compared to $\f{1}{n}$ in some sense. Our main result will involve the 
very weak requirement on the errors that $n(\abs{E_a (n)} + \abs{E_b (n)}) \to 0$. (In fact, we 
only need the weaker condition that $\sum_{j=1}^n (\abs{E_a (j)}+ \abs{E_b (j)})$ is $o(\ln n)$.) 
In discussing the historical context, we will consider stronger assumptions like 
\begin{equation} \lb{6.3}
E_\cdot (n) =\f{\gamma}{n^2} + o\biggl( \f{1}{n^2}\biggr) 
\end{equation}
We will also mention examples where the leading $n^{-1}$ terms are replaced by $(-1)^n n^{-1}$. 

These examples are natural because they are just at the borderline beyond $J-J_0$ trace class 
or $\ul{A}_0 (J) <\infty$ or $\bar A_0 (J)>-\infty$. 

Here is the general picture for these examples. The $(\alpha,\beta)$ plane is divided into four 
regions: 
\begin{SL}
\item[(a)] $\abs{\beta} <-2\alpha$. Szeg\H{o} fails at both $-2$ and $2$. 
\item[(b)] $\abs{\beta} \leq 2\alpha$. Szeg\H{o} holds. 
\item[(c)] $\beta >2\abs{\alpha}$ or $\beta =-2\alpha$ with $\beta >0$. Szeg\H{o} holds at $+2$ 
but fails at $-2$. 
\item[(d)] $\beta <-2\abs{\alpha}$ or $\beta =2\alpha$ with $\beta <0$. Szeg\H{o} holds at $-2$ but 
fails at $+2$. 
\end{SL}

\smallskip
{\it Remarks.} 1. These are only guidelines and the actual result that we can prove requires    
estimates on the errors. 

\smallskip
2. Put more succinctly, Szeg\H{o} holds at $\pm 2$ if and only if $2\alpha \pm\beta \geq 0$. 

\smallskip
3. We need strong hypotheses at the edges of our regions where $\abs{\beta} =2\abs{\alpha}$. For 
example, ``generally" Szeg\H{o} should hold if $\beta =2\alpha >0$, but if $a_n =1 + \alpha n^{-1} 
-(n \ln(n))^{-1}$ and $b_n =2\alpha n^{-1}$, the Szeg\H{o} condition fails (at $-2$), as follows 
from Theorem~\ref{T6.1} below.

\smallskip
Here is the history of these kinds of problems:

\smallskip
(1) Pollaczek \cite{Po1,Po2,Po3} found an explicit class of orthogonal polynomials in the region 
(in our language) $\abs{\beta} <-2\alpha$, one example for each such $(\alpha,\beta)$ with further 
study by Szeg\H{o} \cite{Sz20,Szbk} (but note formula (1.7) in the appendix to Szeg\H{o}'s book 
\cite{Szbk} is wrong --- he uses in that formula the Bateman project normalization of the parameters 
he calls $a,b,$ not the normalization he uses elsewhere). They found that for these polynomials, the 
Szeg\H{o} condition fails.

(2) In \cite{Nev1}, Nevai reported a conjecture of Askey that (with $O(n^{-2})$ errors) 
Szeg\H{o} fails for all $(\alpha,\beta)\neq (0,0)$.

(3) In \cite{AI}, Askey-Ismail found some explicit examples with $b_n \equiv 0$ and $\alpha >0$, and 
noted that the Szeg\H{o} condition holds (!), so they concluded the conjecture needed to be 
modified.

(4) In \cite{DN}, Dombrowski-Nevai proved a general result that Szeg\H{o} holds when $b_n \equiv 0$ 
and $\alpha >0$ with errors of the form \eqref{6.3}.

(5) In \cite{CI}, Charris-Ismail computed the weights for Pollaczek-type examples in the entire $(\alpha, 
\beta)$ plane to the left of the line $\alpha =1$, and considered a class depending on an additional 
parameter, $\lambda$. While they did not note the consequence for the Szeg\H{o} condition, their 
example is consistent with our picture above.

\smallskip
In addition, we note that in \cite{Nev1}, Nevai proved that the Szeg\H{o} condition holds if 
$a_n =1 + (-1)^n\alpha /n + O(n^{-2})$ and $b_n =(-1)^n\beta/n + O(n^{-2})$. 

With regard to this class, here is our result in this paper:

\begin{theorem}\lb{T6.1} Suppose 
\begin{gather}
\sum_{n=1}^\infty (a_n -1)^2 + b_n^2 <\infty \lb{6.4} \\
\limsup_N \biggl(- \sum_{n=1}^N (a_n -1 \pm \tfrac12 b_n) \biggr) =\infty \lb{6.5}
\end{gather}
for either plus or minus. Then the Szeg\H{o} condition fails at $\pm 2$. 
\end{theorem}

\begin{proof} \eqref{6.5} implies that $\bar A_1^\pm (J)=\infty$ so by 
Theorem~\ref{T4.4}{\rm{(ii)}}, $Z_1^\pm (J)=\infty$. 
\end{proof}

{\it Remark.} The same kind of argument lets us also prove the failure of the Szeg\H{o} conditoin 
without assuming \eqref{6.4} by replacing \eqref{6.5} by the slightly stronger condition that 
\begin{equation} \lb{6.5a} 
\limsup_N \biggl( -\sum_{n=1}^N \biggl(\ln (a_n) \pm \f{1}{p}\, b_n \biggr) \biggr) =\infty
\end{equation} 
for some $p>2$. For one can use the step-by-step sum rule for $(1\pm \f{2}{p}\cos (\theta))$. 
\eqref{6.4} is not needed to control errors in $\calE$-sums since they have a definite sign 
near both $+2$ and $-2$, and it is not needed to replace $\ln (a)$ by $a-1$ since \eqref{6.5a} 
has $\ln (a_n)$. 

These considerations yield another interesting result. One can prove Theorem~\ref{T4.1} for the 
weight $w(\theta)=1\pm\f2p\cos(\theta)$ just as we did it for the weight $1$. Since $w(\theta)$ is 
bounded away from zero, the corresponding $Z^\pm$ term is finite if only if $Z$ is. Since $p>2$, 
the corresponding eigenvalue term is finite if and only if $\calE_0$ is. Using 
Theorem~\ref{4.1}(a)--(d) for this $w(\theta)$, we obtain

\begin{theorem}\lb{new} Let $\abs{p}<\f12$ and $\abs{q}<\f12$. 
\begin{SL}
\item[{\rm{(i)}}] If
\[
\limsup_N\biggl(-\sum_{n=1}^N (\ln(a_n)+pb_n)\biggr) >-\infty
\]
and 
\[
\limsup_N\biggl(-\sum_{n=1}^N (\ln(a_n)+qb_n)\biggr)=-\infty
\]
then $Z(J)=\infty$.
\item[{\rm{(ii)}}] If 
\[
\liminf_N \biggl(-\sum_{n=1}^N(\ln(a_n)+pb_n)\biggr)<\infty
\]
and 
\[
\liminf_N \biggl(-\sum_{n=1}^N(\ln(a_n)+qb_n)\biggr)=\infty
\]
then $\calE_0(J)=\infty$.
\end{SL}
\end{theorem}

In particular, if $a_n=1$, $b_n\geq 0$, and $\sum_{n=1}^\infty b_n=\infty$, we have $Z(J)=\infty$ 
and $\calE_0(J)=\infty$. On the other hand, if instead $\sum_{n=1}^\infty b_n<\infty$, then $Z(J)
<\infty$ and $\calE_0(J)<\infty$ (see \cite{KS,HS}).

\begin{corollary} \lb{C6.2} If $a_n,b_n$ are given by \eqref{6.1}, \eqref{6.2} with 
\begin{equation} \lb{6.6}
\lim_{n\to\infty}\,n[\abs{E_a (n)} + \abs{E_b (n)}]=0
\end{equation}
and $2\alpha \pm\beta <0$, then the Szeg\H{o} condition fails at $\pm 2$. 
\end{corollary}

{\it Remarks.} 1. This is intended as separate results for $+$ and for $-$. 

\smallskip
2. All we need is 
\[
\lim_{n\to\infty}\, (\ln N)^{-1} \sum_{n=1}^N (\abs{E_a(n)} + \abs{E_b (n)}) =0
\]
instead of \eqref{6.6}. In particular, trace class errors can be accommodated. 

\begin{proof} If \eqref{6.6} holds, 
\[
\sum_{n=1}^N (a_n -1) \pm \tfrac12 b_n = (\alpha \pm\tfrac12 \beta) \ln N + o(\ln N)
\]
so \eqref{6.5} holds if $2\alpha \pm\beta <0$. 
\end{proof}

As for the complementary region $\abs{\beta}\leq 2\alpha$, one of us has proven (see Zlato\v{s} 
\cite{Zl}) the following: 

\begin{theorem}[Zlato\v{s} \cite{Zl}] \lb{T6.3} Suppose $\abs{\beta}\leq 2\alpha$ and
\begin{align*}
a_n &=1+\alpha n^{-1}+O(n^{-1-\veps}) \\
b_n &=\beta n^{-1}+O(n^{-1-\veps}) 
\end{align*}
for some $\veps>0$. Then the Szeg\H{o} condition holds.
\end{theorem}

{\it Remarks.} 1. This is a corollary of a more general result (see \cite{Zl}).

\smallskip
2. In these cases, $-\sum_{n=1}^N \ln (a_n)$ diverges to $-\infty$. This is only consistent with 
\eqref{4.15} because $\calE_0 (J)=\infty$, that is, the eigenvalue sum diverges and the two 
infinities cancel. 

\smallskip
We can use these examples to illustrate the limits of Theorem~\ref{T4.1}: 
\begin{SL} 
\item[(1)] If $a_n =1$ and $b_n =\f{1}{n}$, then $Z(J)=\infty$ (by Corollary~\ref{C6.2}) while 
$\bar A_0 (J) =\ul{A}_0 (J)<\infty$. Thus $\calE_0 (J)=\infty$. 

\item[(2)] If $a_n=1-\f{1}{n}$, $b_n =0$, then $Z(J)=\infty$ (by Corollary~\ref{C6.2}) $\bar A_0 
(J)=\ul{A}_0 (J) =\infty$, but $\calE_0 (J)<\infty$ since $J$ has no spectrum outside $[-2,2]$.

\item[(3)] If $a_n =1 + \f1{n}$, $b_n =0$, then $Z(J)<\infty$ (by Theorem~\ref{T6.3}), but 
$\bar A_0 (J) =\ul{A}_0 (J) =-\infty$ and so $\calE_0 (J)=\infty$. 
\end{SL}

Finally, we note that Nevai's \cite{Nev1} $(-1)^n/n$ theorem shows that we can have $Z(J)<\infty$, 
$\calE_0(J) <\infty$, and have the sums $\sum a_n$ and/or $\sum b_n$ be only conditionally and not 
absolutely convergent. 

\bigskip


\end{document}